\RequirePackage[2020-02-02]{latexrelease}
\documentclass[a4paper,fleqn]{cas-sc}
\usepackage[numbers]{natbib}
\usepackage{booktabs}
\usepackage{adjustbox}
\usepackage{longtable}
\usepackage{caption}
\usepackage{mathtools}
\usepackage{amssymb}
\usepackage{amsmath}
\usepackage{etoolbox}

\usepackage{subfigure}

\newlength{\myalignwidth}

\def\tsc#1{\csdef{#1}{\textsc{\lowercase{#1}}\xspace}}
\tsc{WGM}
\tsc{QE}
\tsc{EP}
\tsc{PMS}
\tsc{BEC}
\tsc{DE}

\begin{document}
\let\WriteBookmarks\relax
\def\floatpagepagefraction{1}
\def\textpagefraction{.001}
\let\printorcid\relax
\captionsetup[figure]{labelfont={bf},labelformat={default},labelsep=period,name={Fig.}}

% Short title
\shorttitle{An evolutionary game with reputation-based imitation-mutation dynamics}

% Short author
\shortauthors{K. Feng et~al.}

% Main title of the paper
\title [mode = title]{An evolutionary game with reputation-based imitation-mutation dynamics}                      
\author[1]{Kehuan Feng}
\author[1]{Songlin Han}
\author{Minyu Feng}
\fnmark[b]
\cormark[1]
\cortext[cor1]{Corresponding authors: myfeng@swu.edu.cn}
%\ead{myfeng@swu.edu.cn}
\author[3]{Attila Szolnoki}

\affiliation[1]{College of Han Hong, Southwest University, Chongqing, 400715, China.}
\affiliation[2]{College of Artificial Intelligence, Southwest University, Chongqing, 400715, China.}
\affiliation[3]{Institute of Technical Physics and Materials Science, Centre for Energy Research, P.O. Box 49, H-1525 Budapest, Hungary.}

\begin{abstract}
Reputation plays a crucial role in social interactions by affecting the fitness of individuals during an evolutionary process. Previous works have extensively studied the result of imitation dynamics without focusing on potential irrational choices in strategy updates. We now fill this gap and explore the consequence of such kind of randomness, or one may interpret it as an autonomous thinking. In particular, we study how this extended dynamics alters the evolution of cooperation when individual reputation is directly linked to collected payoff, hence providing a general fitness function. For a broadly valid conclusion, our spatial populations cover different types of interaction topologies, including lattices, small-world and scale-free graphs. By means of intensive simulations we can detect substantial increase in cooperation level that shows a reasonable stability in the presence of a notable strategy mutation. 
\end{abstract}


\maketitle

\section{Introduction}
\label{intro}
The emergence of cooperation in a group of self-interest actors is an intensively studied problem among researchers originated from various academic fields~\cite{perc_pr17,nowak_06}, including microbiology, ecology, economics, and social sciences~\cite{kreft_mb04,chica_srep19,axelrod_84}. In natural world, both in animal kingdom and human societies, there are several exotic examples of cooperation, such as food-sharing among vampire bats~\cite{wilkinson_n84}, or 'marine snow'~\cite{datta_cb14}. But cooperation is also essential to address vital challenges of climate change or environmental protection~\cite{milinski_pnas08,szolnoki_plrev14,he_jl_pla23,chen_xj_pcb18}. Despite of the collective benefit for mutual cooperation, defection is still tempting in a social dilemma situation because it offers the highest individual payoff for a defector~\cite{axelrod_s81}.

Evolutionary game theory~\cite{maynard_82,sigmund_10} was proposed to address this problem where the various form of conflict is described by some frequently studied games, including prisoner's dilemma game (PDG) \cite{rapoport_70,zhang_y_amc24,YAO.2,kojo_csf24, Rong.1}, snowdrift game (SDG) \cite{li_k_csf21,feng_my_csf23, Zeng}, stag hunt game (SHG) \cite{skyrms_04,deng_ys_epjb22} and public goods game (PGG) \cite{YAO.1, szolnoki_epl10}. As a key finding, the spatial setting of participants is an essential element to reach decent cooperation level even at harsh conditions when defection is attractive otherwise. This observation launched a bloom of research activity where different interaction graphs were studied systematically. Starting from the simplest lattice topology, the consequence of random graph~\cite{szolnoki_epl09,duran_pd05,vukov_pre06}, small-world networks~\cite{kim_bj_pre02,wu_zx_pre05,bin_pa22}, scale-free networks~\cite{santos_prl05,szolnoki_pa08}, or even more complex interdependent and multiplex graphs were revealed~\cite{wang_z_epl12,li_q_e22,wang_z_srep13, Rong.2}.

Beside spatial setting, alternative cooperation supporting elements are also identified, like various ways of reciprocity, reputation \cite{yang_hx_pa19,quan_j_pa21}, punishment \cite{helbing_ploscb10,brandt_prsb03,lee_hw_amc22}, exclusion~\cite{liu_lj_rspa22,szolnoki_pre17,quan_j_jsm22,sun_xp_pla23}, reward~\cite{hua_sj_eswa24}, persistence \cite{zhang_lm_epl19,liu_dn_pa19}, etc. Furthermore, to go beyond the simplest approach of binary strategy choice of unconditional strategies, more sophisticated multi-strategy models enrich the diversity of individual actions. Just to mention a few, tit-for-tat~\cite{nowak_n93,sasidevan_srep16} or win-stay-lose-shift strategy updates~\cite{fu_mj_ijmpc18}, but this research path also includes interactive diversity where the same player behaves differently toward different neighbors~\cite{su_q_njp16}.

We here focus on reputation, as a fruit of the most complex and distinctive activities in human society, that extensively influences the individual and group choices during interactions~\cite{quan_j_jsm22,M.Feng,wei_x_epjb21,bin_l_amc23}. Bad reputation, for instance, has serious consequences on individual success. The involved members are not popular in a society, but people with low credit scores also have difficulties to apply for a loan in banks. In today's increasingly digitized societies, personal information has become more transparent, allowing people to access information about the reputation and social circles of their associates via social media, which allows people to gather reliable information about the willingness of cooperation of others. 

The other key element of our model is the microscopic dynamics, the way how a player changes strategy. A reasonable choice is imitation which is a prevalent instinct observed among animals, and, of course, humans are no exception \cite{grujic_rsos22}. Individuals often imitate those who surpass them in certain aspects, aiming to improve themselves or reap greater rewards. In a particular example, students in modern societies diligently imitate the learning methods and even lifestyle habits of high-achievers, aspiring to achieve academic success, just like their role models \cite{rizolatti_arn04}. In a recent related study, Zhang {\it et al.} investigated the influence of asymmetric comparison of fitness based on reputation \cite{zhang_zp_pa22}. Undoubtedly, imitation serves as a common and representative rule for strategy updating, most of the literature primarily focus on this aspect. However, it is important to acknowledge that human society is replete with ``irrational choices'' or ``sudden acts'' where individuals do not rely solely on a single criterion to make strategic decisions\cite{irrational.1, irrational.2}. Hence, it remains unexplored how additional updating protocol, in parallel with the leading imitation process, affects the collective behavior.

In previous works where the influence of reputation on cooperation was studied, researchers mostly focused on the imitation rule without assuming additional microscopic effects. However, some earlier papers already highlighted that individuals possess a certain probability of mutation or ``exploration rate'' besides inheriting the strategies of the parent generation, to express the initiative and uncertainty in strategy selection \cite{traulsen_pnas09,santos_srep16,erovenko_prsa19}. Motivated by these observations, we propose a new model to discuss the influence of reputation on the evolution of cooperation in spatial populations. Initially, the reputation value of each individual is set to be a starting value which is then updated after each game with neighbors. During microscopic dynamics, we combine the imitation process with a random individual strategy mutation, which helps us to explore the potential consequences of irrational decision-making and psychological factors in collective behavior. 

The remaining of our paper is organized as follows. Section 2 describes our models and evolutionary dynamics in detail. In Section 3, the simulation results showcase the influence of the imitation-mutation strategy update rule on the evolution of cooperation compared to the traditional imitation rule. Section 4 gives comprehensive conclusions drawn from this work.

\section{Imitation-mutation model}

Previous studies focusing on reputation mostly considered imitation as an individual strategy updating rule (referred as IM in the following). To generalize and extend these microscopic dynamics, we propose an imitation-mutation strategy update rule (referred as IM-MUTA). This protocol differs significantly from previous cases, as the introduced mutation rate to some extent represents irrational choices made by humans. In this section, we mainly summarize the details of our game model based on individuals with reputation mechanisms that is considered in individual’s payoff which mostly determines strategy-updating process. First, we summarize those games where we study social dilemma situations.

\subsection{Social dilemma for PDG and SDG}
In the PDG, each individual chooses between cooperation (C) or defection (D). In case of mutual cooperation, each individual receives a payoff of $R$. For mutual defection, both players are punished by payoff $P$. When $C$ and $D$ players meet, the former receives a payoff of $S$, while the latter enjoys the highest payoff of $T$. The rank $T > R > P > S$ defines a PDG, where defection is the better individual strategy independently of the other's choice.  In the context of SDG, which uses the same strategies and payoff parametrization, the main difference is the $T > R > S > P$ rank of payoff elements. This subtle alteration in the payoff structure leads to an alternative Nash-equilibrium formed by a C-D pair. In essence, a rational player adopts the opposite strategic stance represented by the partner in the realm of SDG.

By following previous works we fix $R=1$ and $P=0$ parameter values and the remaining two free payoff elements characterize the games. Without losing the essence of dilemmas we further reduce the number of free parameters and we determine both $T$ and $S$ values by a single parameter. In particular, for PDG we use $T=1+r_1$ and $S=-r_1$, while for SDG $T=1+r_2$ and $S=1-r_2$. Summing up, the corresponding payoff matrix is 
\begin{equation}\label{eq1}
M_1=
 \left(
\begin{array}{cc}
     1&-r_1  \\
     1+r_1&0 
\end{array}
\right)   
\end{equation}
for PDG, while the $M_2$ matrix for SDG is
\begin{equation}\label{eq2}
M_2=
 \left(
\begin{array}{cc}
     1&1-r_2  \\
     1+r_2&0 
\end{array}
\right) .
\end{equation}
In both cases parameters $r_1$ and $r_2$ remain in $0<r_1<1$, $0<r_2<1$ interval. For both games the higher the $r_1$ or $r_2$ the greater the temptation to defect.

All individuals are arranged on a network where they interact with their neighbors and collect an accumulated payoff $P_i(t)$ at time $t$ according to a payoff matrix described above. 

\subsection{Evolutionary dynamics}

During an elementary step, we update not only the $s_i$ strategy of individual $i$ but also its $R_i$ reputation. Suppose that $R_i = 0$ at $t=0$ time. Later this value may change depending on the actual strategy represented by the actor in the latest step. In general, $R_i$ is increased by a value $\alpha$ if player $i$ becomes a cooperator and it is decreased by the same amount in the reversed case. The time evolution of individual reputation can be summarized as
\begin{equation}\label{eq3}
    R_i(t)=
    \begin{cases}
    R_i(t-1)+\alpha, s_i(t)=C\\
    R_i(t-1)-\alpha, s_i(t)=D \,.
\end{cases}
\end{equation}
Importantly, the above defined $R_i$ value always remains in the $0\le R_i \le 2$ interval. Another key assumption of our model is the reputation value affects the fitness level of players directly, hence it has straightforward impact on strategy update. In particular, the total fitness of player $i$ at time $t$ is defined by
\begin{equation}\label{eq4} 
F_i(t)=R_i(t) \times P_i(t)\,,
\end{equation}
where the accumulated $P_i(t)$ payoff value is calculated from the interactions with neighbors according to the payoff matrix defined by Eq.~(\ref{eq1}) or Eq.~(\ref{eq2}).

\begin{figure}
	\centering
	\includegraphics[scale=0.2]{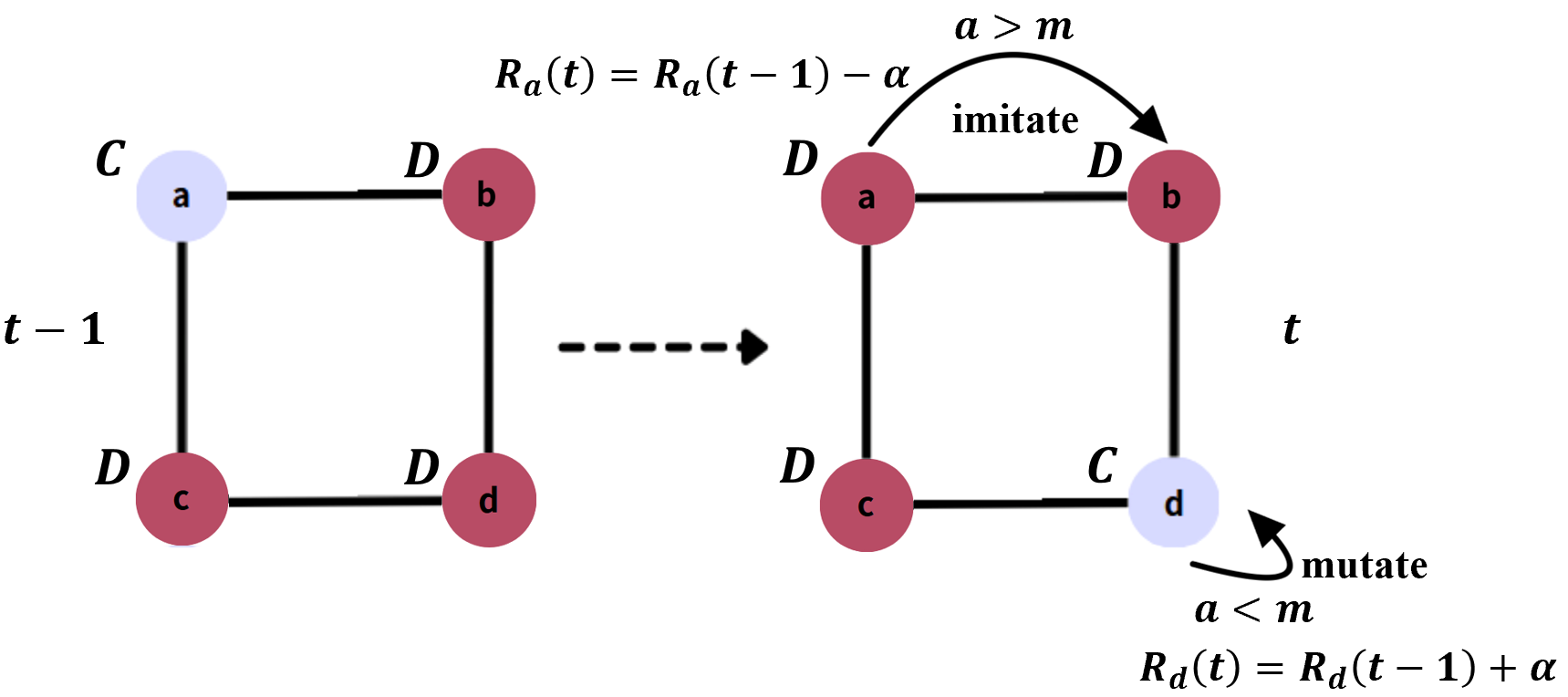}
	\caption{\textbf{Schematic diagram of the imitation-mutation process and reputation mechanism.} At $t-1$ time, player `a` is a cooperator while others, player `b`, `c`, and `d` are defectors. In the next step player `a` imitates the strategy of player `b` with probability $1-m$, while player `d` chooses random mutation with probability $m$. Meanwhile, the reputation index is updated for both players as indicated.}
	\label{FIG:1}
\end{figure}

The extended imitation-mutation dynamics is defined in the following way. With probability $m$ the selected player $i$ will change its current strategy randomly independently of the state of the neighborhood, otherwise, with probability $1-m$, the player follows the standard imitation protocol and imitates the strategy of a randomly chosen neighbor $j$ with probability $H_i$. This probability, as we noted, depends on the extended fitness values of involved partners:
\begin{equation}\label{eq5}
    H_i(t)=\frac{1}{1+\exp[(F_i(t)-F_j(t))/K]}\,.
\end{equation}
Here $F_i(t)$, $F_j(t)$ are the fitness of individual $i$ and $j$, respectively and parameter $K$ represents the noise level of imitation. In the following, we use $K=0.7$ value for the imitation process.

To gain a more comprehensive view of the consequence of the extended dynamics and fitness function, we consider two different social games, as described above, and apply various interaction topologies. In particular, to describe the interactions between players we use square-lattice graphs with periodic boundary conditions, small-world (WS), and scale-free (BA) networks. In the last two graphs, the average degree is $⟨k⟩ = 4$ and $⟨k⟩ = 6$, respectively. In all cases, we consider a population of $N = 10000$ individuals.

According to the standard protocol, we launch the evolution from a random initial state where players are cooperators or defectors with equal weight. In each iteration step on average, all players have a chance to update their strategy and reputation index. When reaching the final stationary state after 100000 steps we calculate different quantities, like the fraction of cooperators or the average fitness level. For reliable statistics, each experiment was repeated 10 times. 

Figure~\ref{FIG:1} summarizes the elementary steps of our proposed model. As it is illustrated, players in the network have two ways to update their strategies. While player `a` applies imitation and adopts the defector strategy of neighboring `b`, player `d` updates its current strategy randomly, independently of the status of neighbors. The latter happens with probability $m$, while the former with probability $1-m$. In parallel, the reputation index of these players is also changed, increased, or decreased by $\alpha$ value, according to their latest strategy.

\section{Results}

In this section, we first present the results obtained for PDG. The $f_C$ stationary portion of cooperators is measured at different values of $r_1$ which represent different strengths of the prisoner's dilemma. Furthermore, we also check the impact of parameters $\alpha$ and $m$ on the $f_C$ level. The system behavior is analyzed for different graph structures, as indicated. After we extend our study by considering SDG situation. For all experiments, in the initial state, the proportion of cooperators and defectors in the network is both 50\%, which means that half of the individuals in the network are cooperators, and the other half are defectors. 

\subsection{Evolution of cooperation in PDG under IM-MUTA dynamics}

\begin{figure*}[htpb]
	\centering
	\subfigure[SL with IM]{
		\includegraphics[scale=0.15]{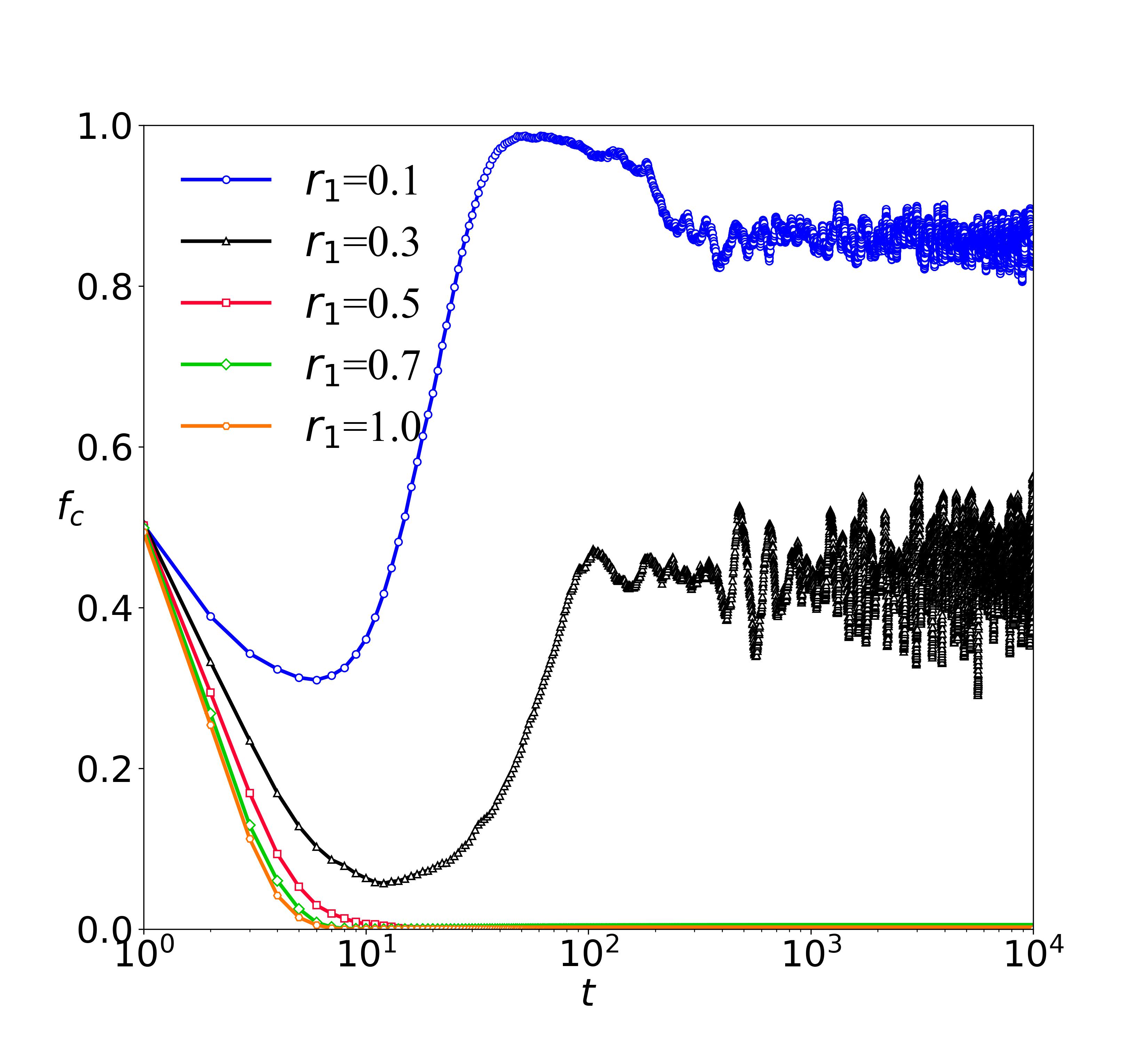}
  }
	\subfigure[WS with IM]{
		\includegraphics[scale=0.15]{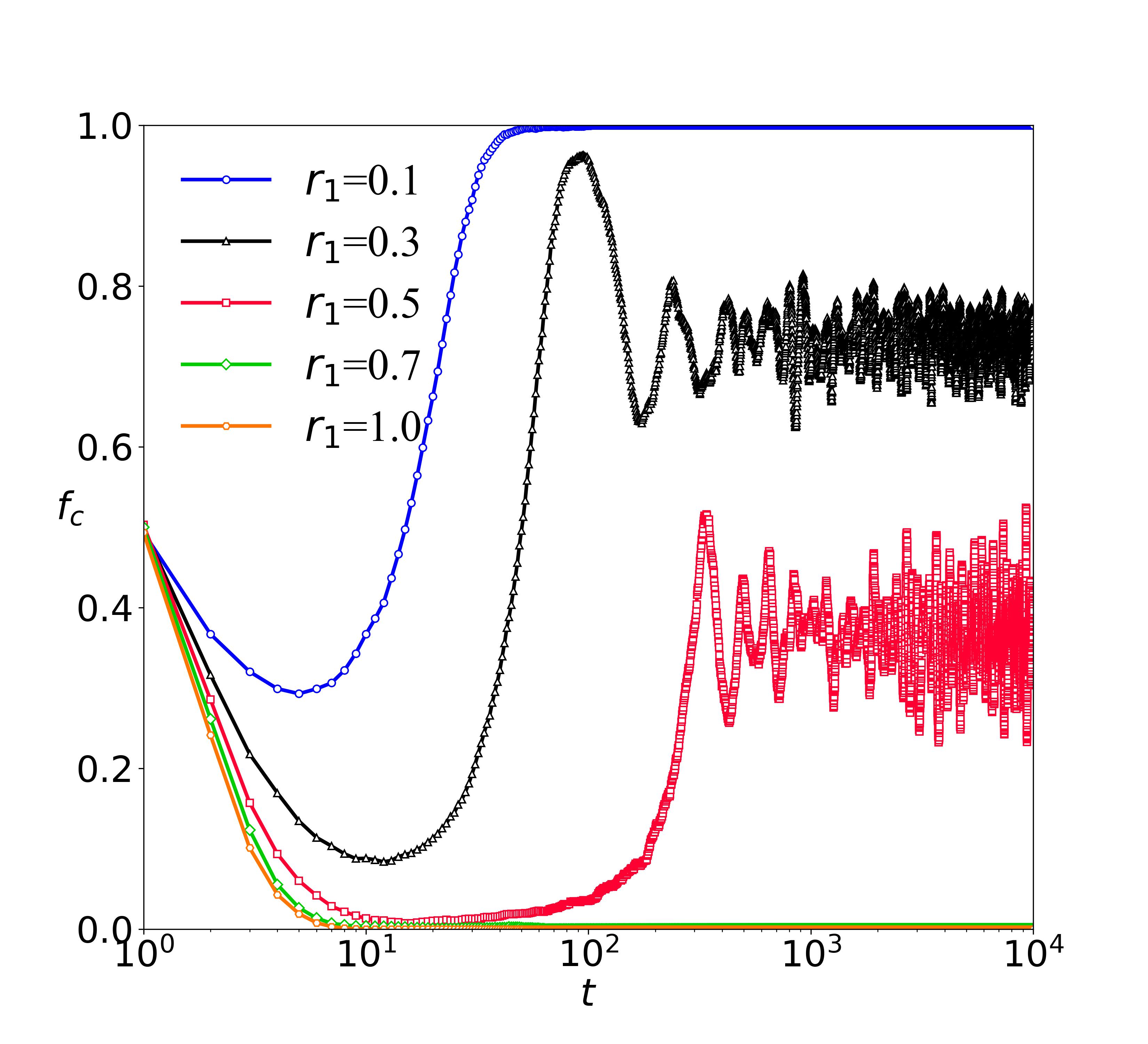}
  }
        \subfigure[BA with IM]{
		\includegraphics[scale=0.15]{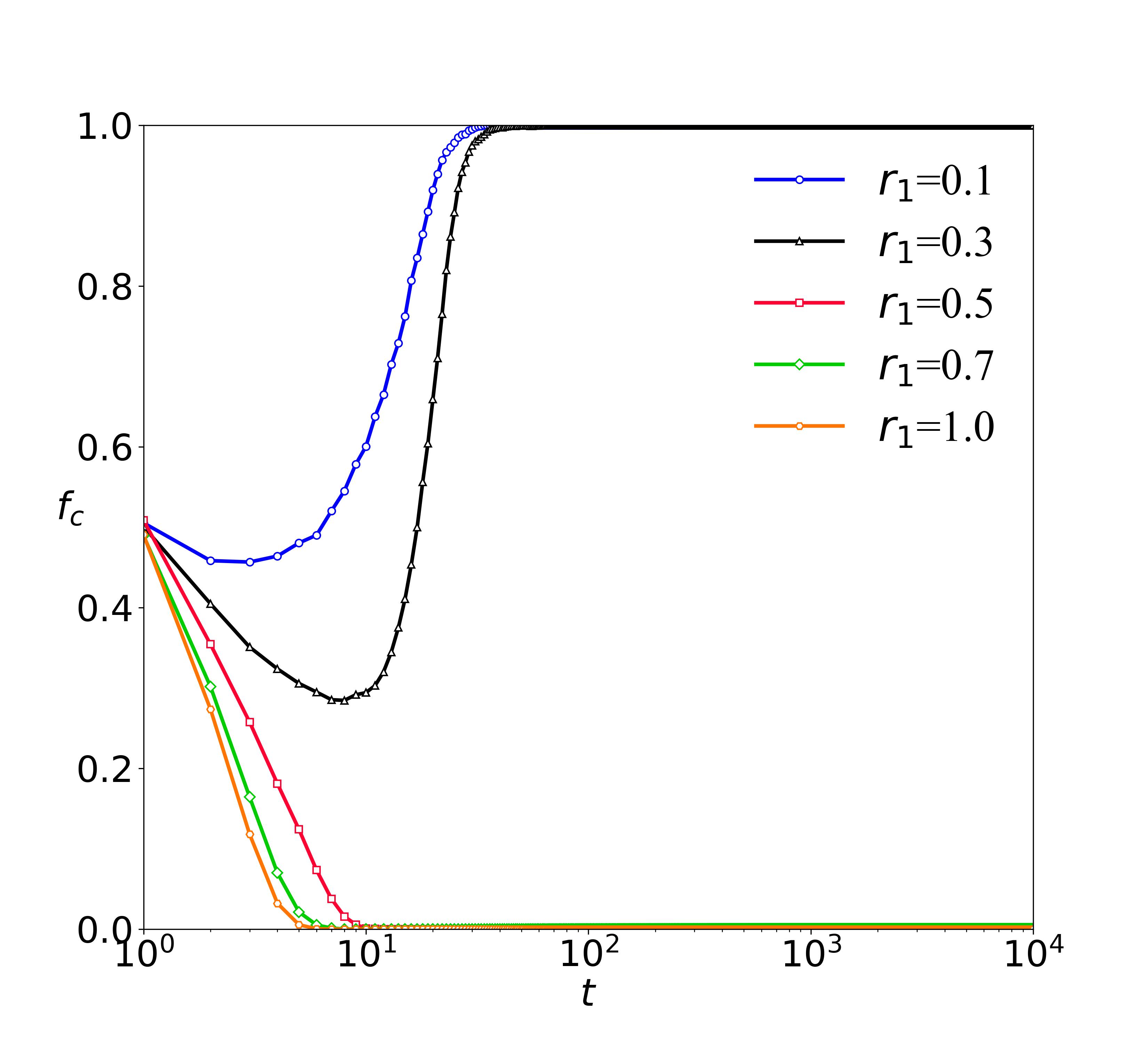}
  }
        \subfigure[SL with IM-MUTA]{
		\includegraphics[scale=0.15]{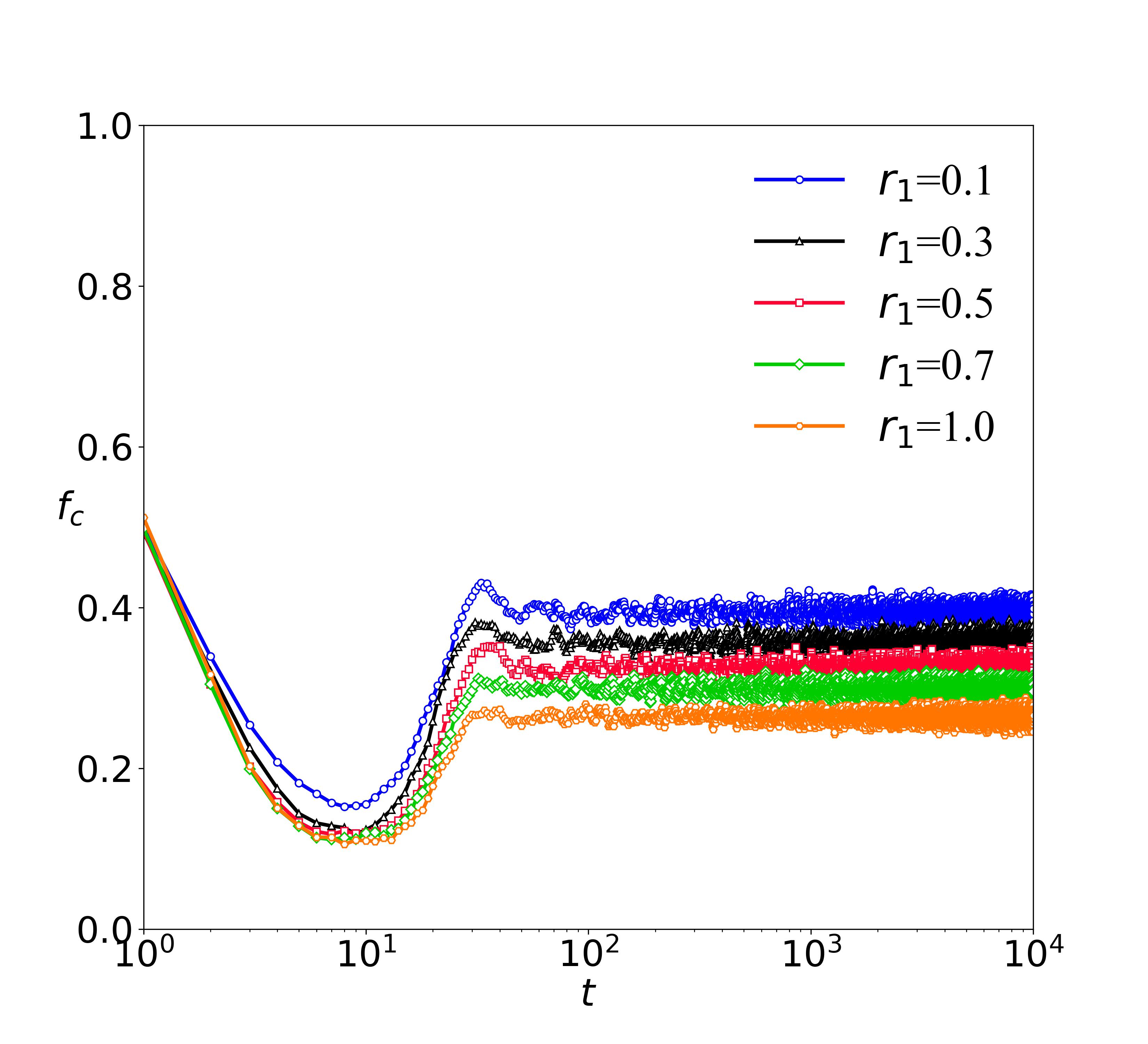}
  }
        \subfigure[WS with IM-MUTA]{
		\includegraphics[scale=0.15]{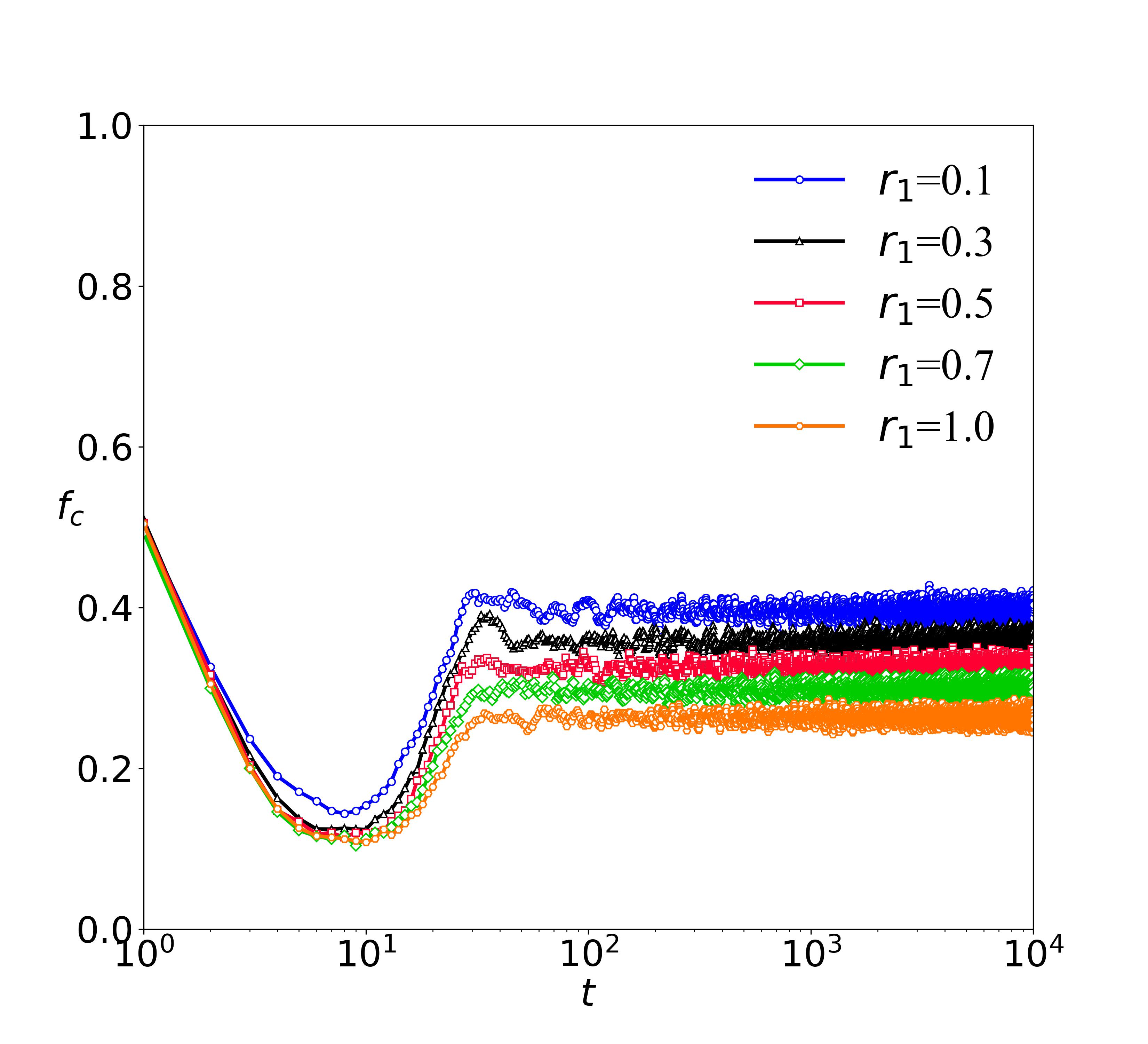}
  }
        \subfigure[BA with IM-MUTA]{
		\includegraphics[scale=0.15]{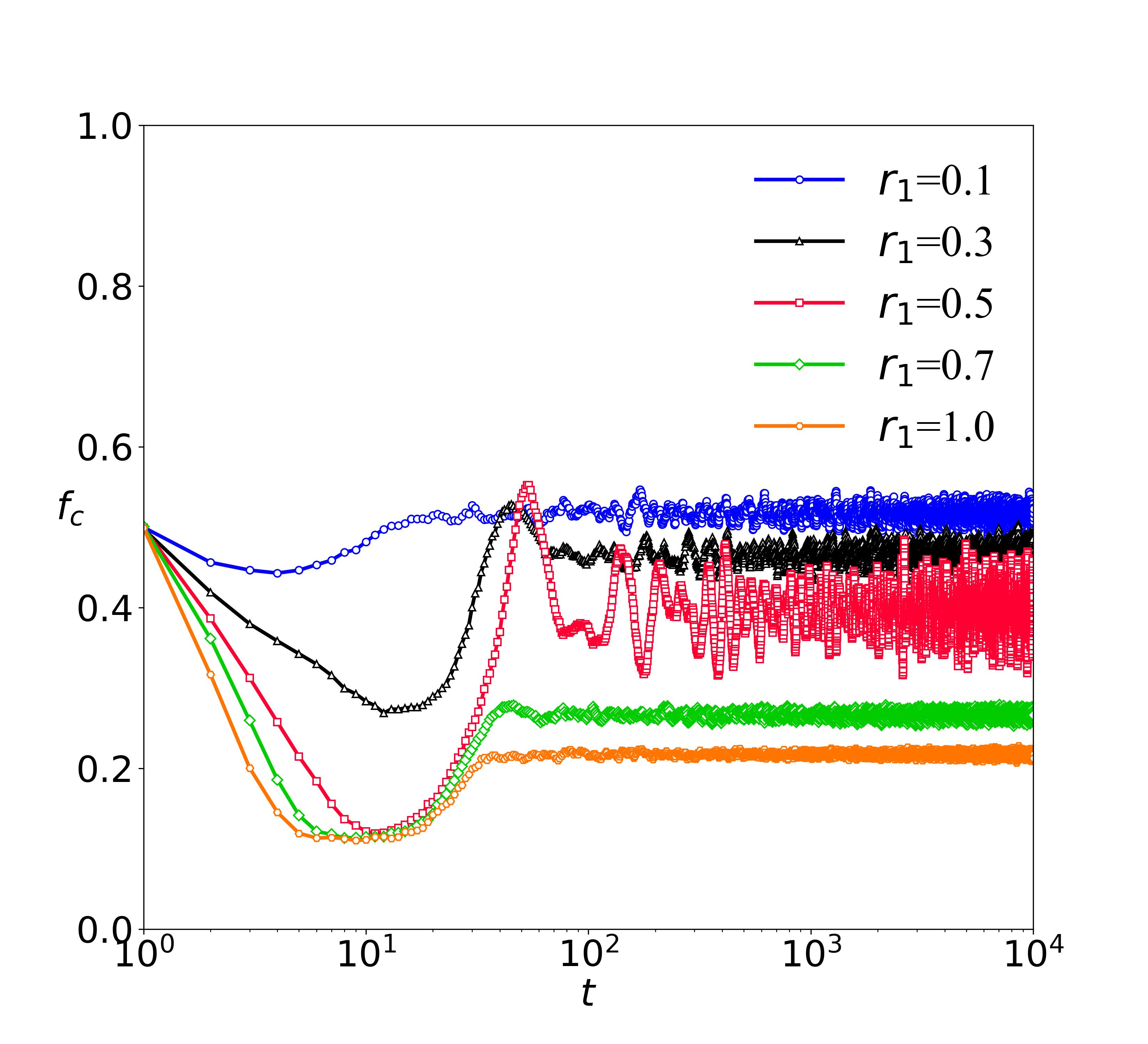}
  }
    %\captionsetup{justification=centerfirst}
	\caption{\textbf{The evolution of cooperation in PDG for different dynamics and topology.} Curves show the time evolution of $f_C$ starting from a random initial state for different $r_1$ values as indicated in the legend. The top row illustrates the evolution of the traditional model where players always follow imitation (IM) during strategy updates. Panel~(a) to (c) represent different interaction graphs, as shown in the labels. As a comparison, the bottom low depicts those cases where the extended imitation-mutation (IM-MUTA) strategy update is applied. Other parameters are $\alpha=0.05$ and $m=0.2$. Note that we used a semi-log plot to stress the time dependence faithfully. The time evolution of $f_C$, ``first down, later up'', demonstrates how network reciprocity works.}
\label{FIG:2}
\end{figure*}

To gain a first impression about the consequence of extended updating dynamics we first present the time evolution of cooperation level obtained for different parameter values and conditions of topology. An overview can be seen in Fig.~\ref{FIG:2}, where the typical evolution is shown starting from a random initial state. In each case we used a broad variety of $r_1$ parameters, ranging from 0.1 to 1.0, to span both weak and strong dilemma situations. To identify the robust system behavior we applied various interaction topologies, such as lattice, panel~(a) and (d), small-world random graph, panel~(b) and (e), and last highly heterogeneous scale-free network, shown in panel~(c) and (f).

Most importantly, we compare the cases where the traditional and the extended updating dynamics are applied. The first row of Fig.~\ref{FIG:2} shows the evolution when players always follow the imitation protocol. To reveal the details of time evolution we use semi-log plots for all panels shown in this figure. It is a common feature for all networks that cooperators cannot survive if $r_1$ is high enough because, in this parameter region, the advantage of defection is overwhelming. For small $r_1$ values, however, they can survive. It is again a generally valid feature that in the latter case there are ``first down, later up`` dynamics in the cooperation level. This is a well-known indication of how network reciprocity works~\cite{perc_pre08b,szolnoki_epjb09}. Namely, cooperators are sensitive in a randomized initial state, but surviving cooperators can form a compact domain and this domain can eventually grow in the sea of defectors. The only difference between different graphs is cooperators and defectors can coexist at appropriate values of $r_1$ on lattices or in random graphs, while there is a sharp transition between dominant states in a highly heterogeneous scale-free graph.

When mutation, or exploratory updating dynamics is also present, shown in the bottom row of Fig.~\ref{FIG:2}, the time dynamics change significantly. Here we applied $\alpha=0.05$ and $m=0.2$ parameter values. The most striking difference is the extended dynamics provide a ``safer'' trajectory for cooperators. More precisely, they can survive for all $r_1$ values in all interaction graphs. We should stress that this behavior is not a straightforward consequence of randomized strategy update because the stationary value of $f_C$ is always higher than the one we would expect based on the value of $m$. Notably, the non-monotonous feature of time evolution can be detected again, signaling that network reciprocity is still working. Therefore we can conclude that there is synergy between imitation and exploratory mutation-based strategy updates which provide a more efficient protocol for cooperation even in very harsh conditions. The other generally valid observation is the usage of the IM-MUTA protocol seems to mitigate some of the effects associated with network topology, leading to a convergence in the cooperative evolution trends across all networks.

\subsection{The effect of $\alpha$ on cooperation level}

\begin{figure}
	\centering
		\includegraphics[scale=.55]{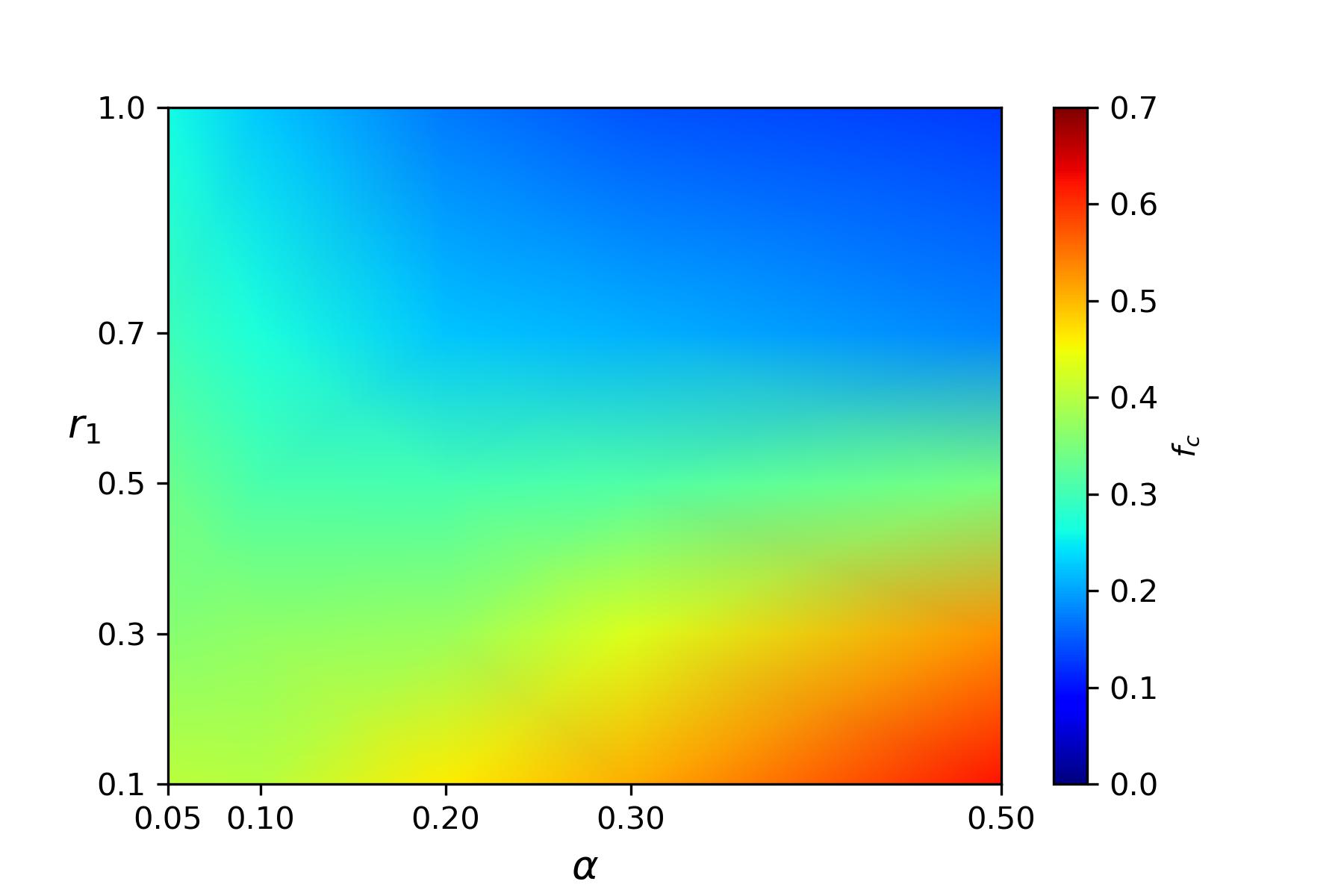}
    %\captionsetup{justification=centerfirst}
	\caption{\textbf{Stationary fraction of cooperators on $r_1 - \alpha$ parameter plane.} The results are obtained for WS interaction graph by using $m=0.2$. The impact of parameter $\alpha$ is ambiguous, while for low $r_1$ values, it is better to use large $\alpha$ steps during the reputation update, for larger $r_1$ values smaller $\alpha$ provide a larger cooperation level. The largest cooperation level can be reached in the low $r_1$ - large $\alpha$ corner of the parameter plane. Note that cooperation level cannot reach 1 even for large $r_1$ due to mutation mechnaism.}
	\label{FIG:3}
\end{figure}

The next crucial point is to explore how parameter $\alpha$ affects our observations. More precisely, to clarify if there is any significant consequence how drastically we change the reputation index of players during an elementary step. One can argue that the answer may largely depend on the dilemma strength, namely on the value of $r_1$. To answer this question properly, in Fig.~\ref{FIG:3} we present the stationary cooperation level on the $r_1 - \alpha$ parameter plane. We here use WS small-world graph which practically represents all significant system behavior, as we previously illustrated in Fig.~\ref{FIG:2}. 

As we discussed in the model definition, parameter $\alpha$ characterizes the way how intensively we adjust the reputation index due to the strategy change of an actor. When the value of $\alpha$ is small then the cooperator act is rewarded gently, while large $\alpha$ values represent strong direct support for cooperation. Based on this interpretation we may expect larger improvement for larger $\alpha$ values, but this expectation is just partly justified. As Fig.~\ref{FIG:3} highlights, we can reach significant improvement in cooperation for higher $\alpha$ values, but only for smaller $r_1$ values. This picture becomes the opposite if $r_1$ exceeds $r_1=0.5$ value. Above this threshold level in the dilemma, it is detrimental to apply large steps when reputation is updated. Instead, we can reach the highest cooperation level when $\alpha$ is small.

\begin{figure}
	\centering
		\includegraphics[scale=0.2]{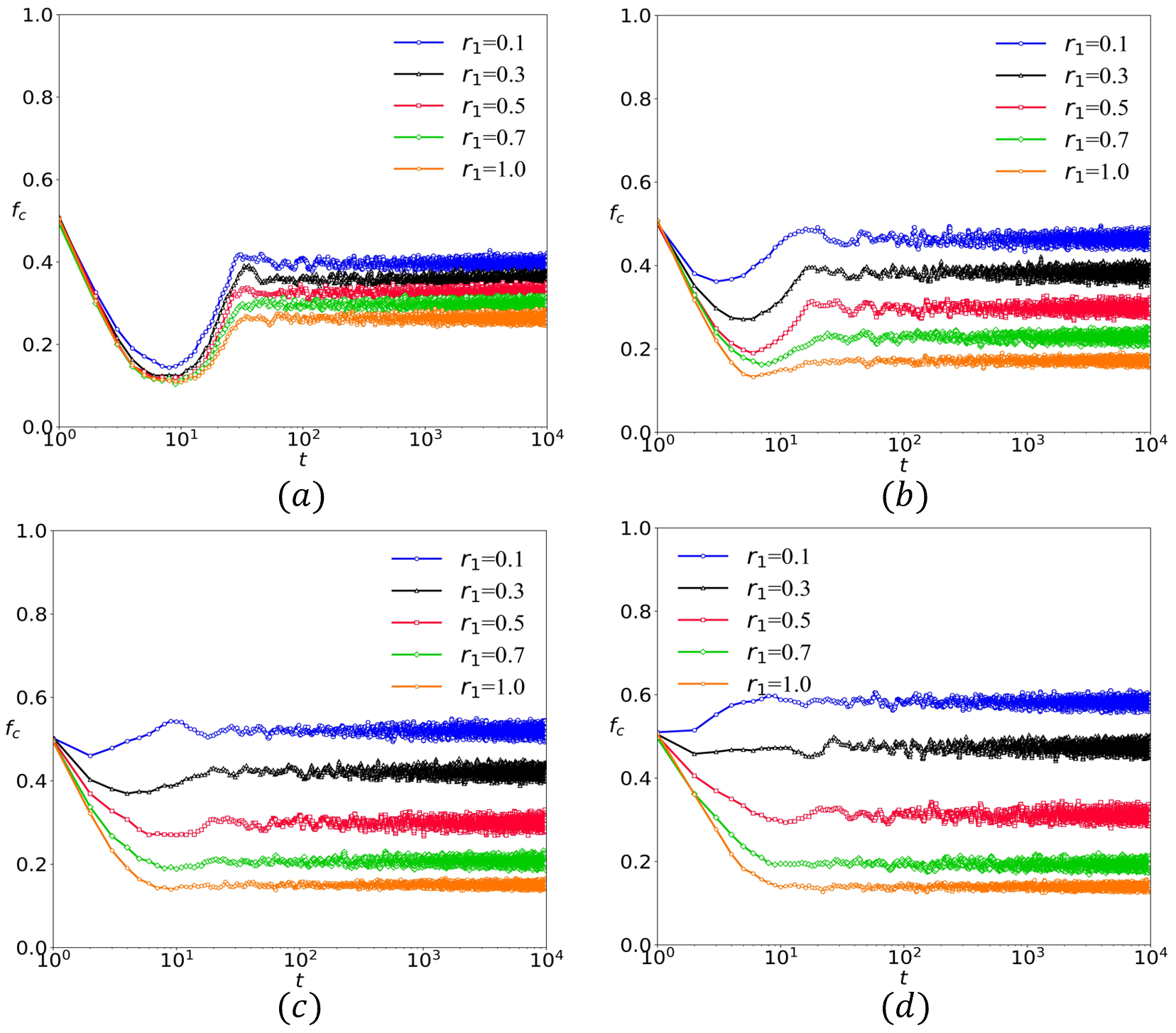}
    %\captionsetup{justification=centerfirst}
	\caption{\textbf{Time evolution of cooperation on WS graph obtained for different $\alpha$ values.} Panel~(a) to (d) respectively shows the cases for $\alpha=0.05$, $0.2$, $0.3$, and $0.4$. In all cases, $m=0.2$ was fixed. The applied $r_1$ values are indicated in the legend for each panel. The comparison of trajectories suggests that the impact of network reciprocity is practically diminished if $\alpha$ exceeds $0.3$.}
	\label{FIG:4}
\end{figure}

The latter phenomenon can be explained by the fact that, under the highest temptation condition ($r_1=1.0$), a higher reputation update coefficient ($\alpha$) rapidly diminishes individuals' reputations in the network, rendering the strategy update based on Eq.~\ref{eq4} less effective. Consequently, cooperation is primarily sustained through strategy updates based on Eq.~\ref{eq5}. Furthermore, when examining the change in the cooperation frequency at $\alpha=1$, we observed consistent behavior with $\alpha=0.5$. Namely, under high temptation values, the cooperation frequency curves exhibit a rapid fall that stabilizes at lower values without exhibiting any local minima. However, when $\alpha$ is less than or equal to $0.3$, in the context of high temptation, although the cooperation frequency also decreases, after some early iterations (around $10^1$), the cooperation frequency curve shows a slight rebound and stabilizes. In this case, the overall cooperation curve exhibits a local minimum. In Fig.~\ref{FIG:4}(a), where $\alpha=0.4$, the trends of the cooperation differ from those shown in Figs.~\ref{FIG:4}(a), (b), and (c). In Figs.~\ref{FIG:4}(a) and (b), it is clear that the curves initially decline, then rise, and finally saturate after a slight decrease. In Fig.~\ref{FIG:4}(c), although the trends of the curves are not as pronounced as in panels ~(a) and (b), they are still qualitatively similar. However, when $\alpha=0.4$, in Fig.~\ref{FIG:4}(d), the curves initially rise and then saturate for smaller $r_1$ values. As $r_1$ increases, the cooperation level decays first and then saturates without showing a local temporary minimum. Based on Fig.~\ref{FIG:3} and Fig.~\ref{FIG:4}, we can conclude that there is no clear connection between cooperation and evolving reputation mechanism because the consequence of $\alpha$ depends sensitively on the temptation level.

\begin{figure}[htbp]
\centering
        \subfigure[]{
		\includegraphics[scale=0.12]{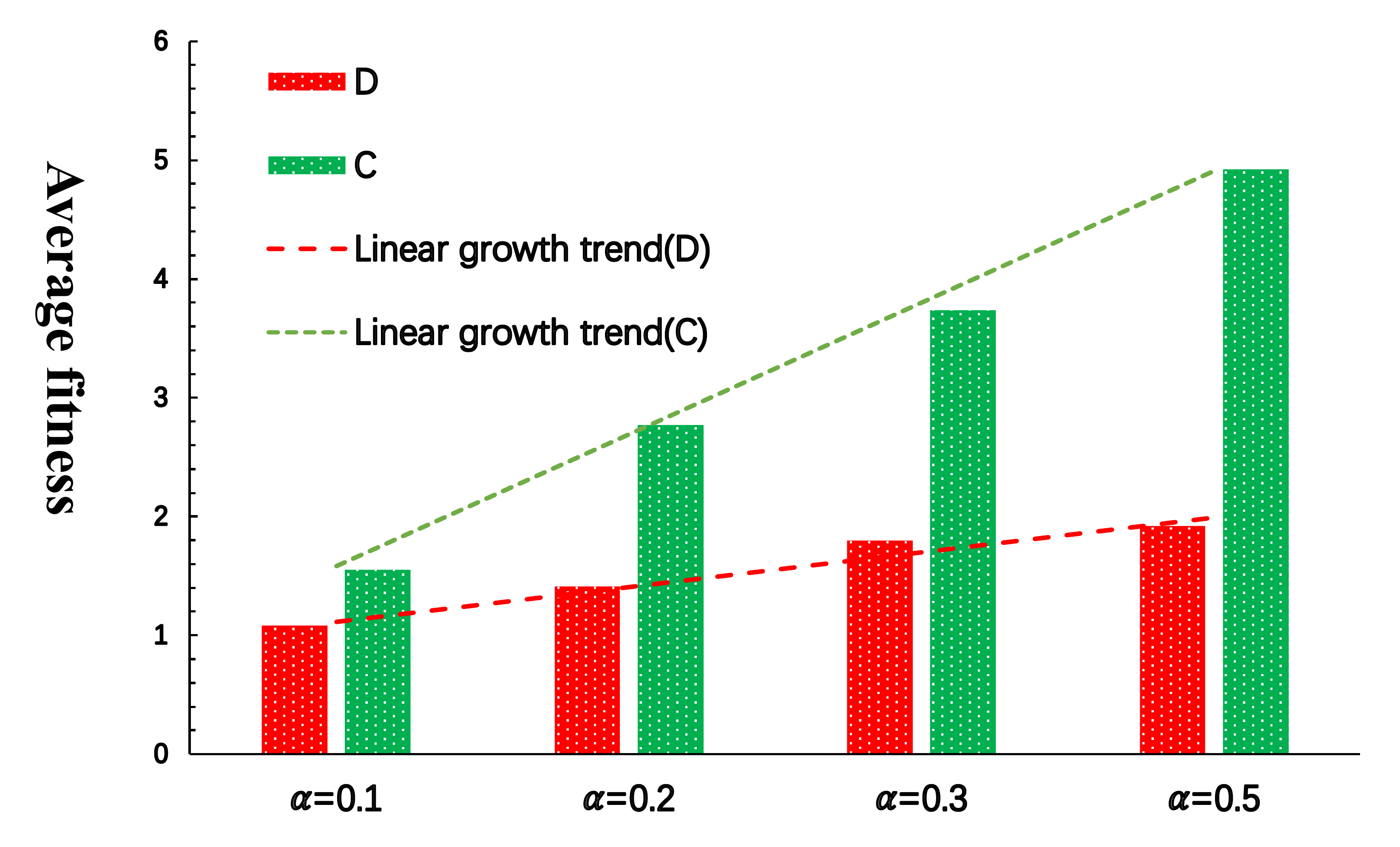}}
	\subfigure[]{
		\includegraphics[scale=0.12]{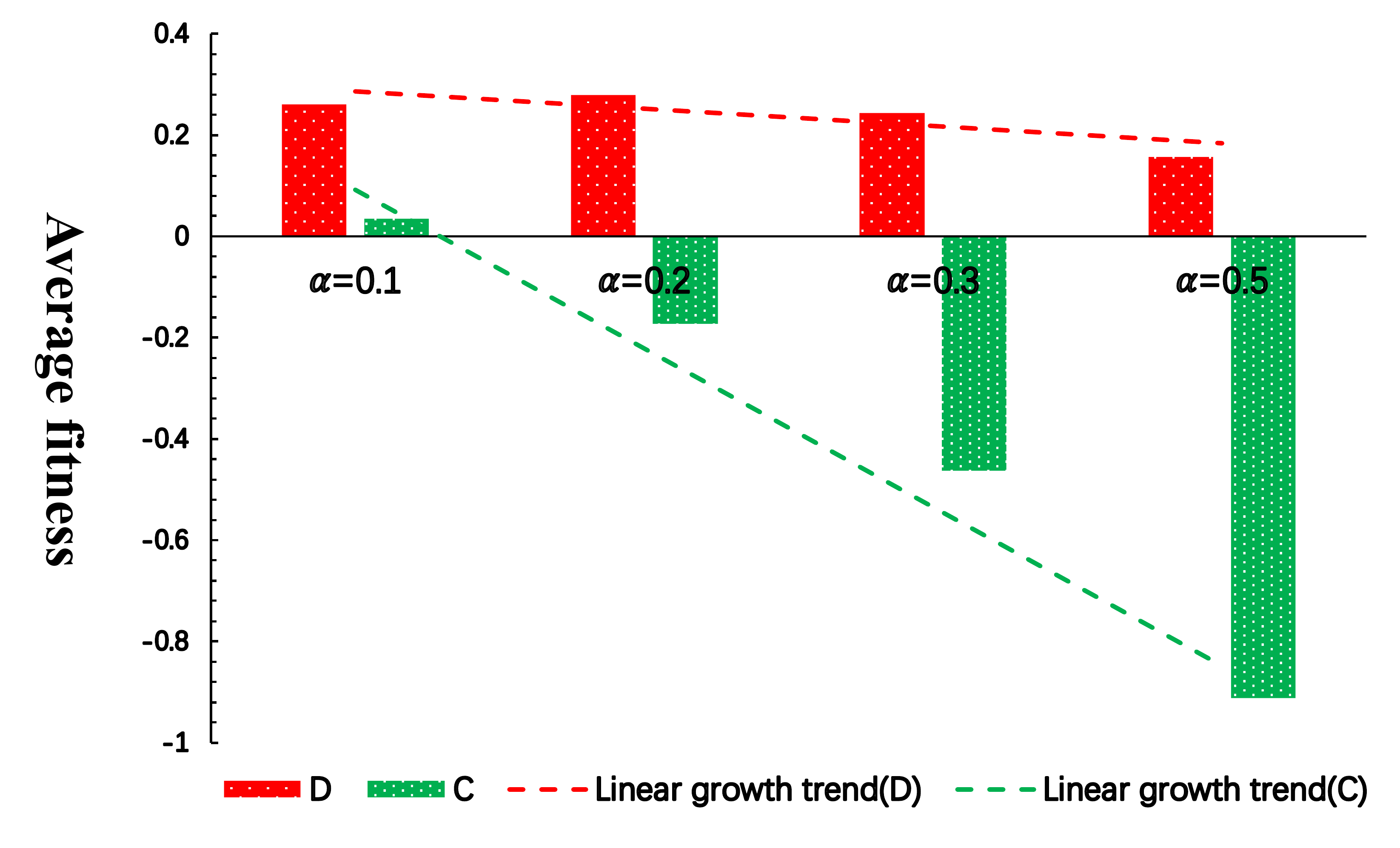}} 
    %\captionsetup{justification=centerfirst}
    \caption{\textbf{The average fitness of strategies at different temptation values in PDG on WS graph.} With IM-MUTA, Panel~(a) depicts how fitness values change for different $\alpha$ for low temptation at $r_1=0.1$. Panel~(b) shows the same quantities for $r_1=0.7$ which represents high temptation. The fitted lines indicate clearly that at low temptation cooperators benefit more if we increase the reputation step $\alpha$. At large temptation, however, cooperators suffer more from the usage of larger $\alpha$.}
    \label{FIG：5}
\end{figure}

To gain a deeper understanding of the counter-intuitive phenomenon discussed above we also measured the average fitness of competing strategies. To reveal the difference resulting in diverse system behavior we selected two representative $r_1$ values from the low- and large-temptation regions. Their comparison can be seen in Fig.~\ref{FIG：5}. The first panel summarizes the low temptation case, obtained at $r_1=0.1$ for different $\alpha$ values, as indicated. The increase of $\alpha$ leads to an increase in the average fitness both for cooperators and defectors, but in a different way. In particular, the increment due to large $\alpha$ is significantly larger for cooperators, hence they can benefit more from the intensive change of reputation index. As a result, the general cooperation level grows by increasing $\alpha$ in the low-temptation region.
Panel~(b) illustrates what is happening when the temptation level is significant. In this case, both defectors and cooperators gain less if we increase $\alpha$. But the change in the general fitness of cooperators is significantly larger, compared to the change for defectors, where the decline is just moderate. Accordingly, the total change is negative, leading to a low-cooperation state for the system. At this point, the drastic change in reputation becomes a burden, causing cooperators to plunge into an abyss and triggering a chain reaction that leads to a state mostly dominated by defection. Such situations are common in real life, e.g., when a large-scale bank experiences a run, it can shake the entire country's financial industry and subsequently trigger a credit crisis throughout society.

\subsection{ Effect of mutation rate $m$ on the cooperation}\

\begin{figure}
    \centering
    %\captionsetup{justification=centerfirst}
    \includegraphics[scale=.2]{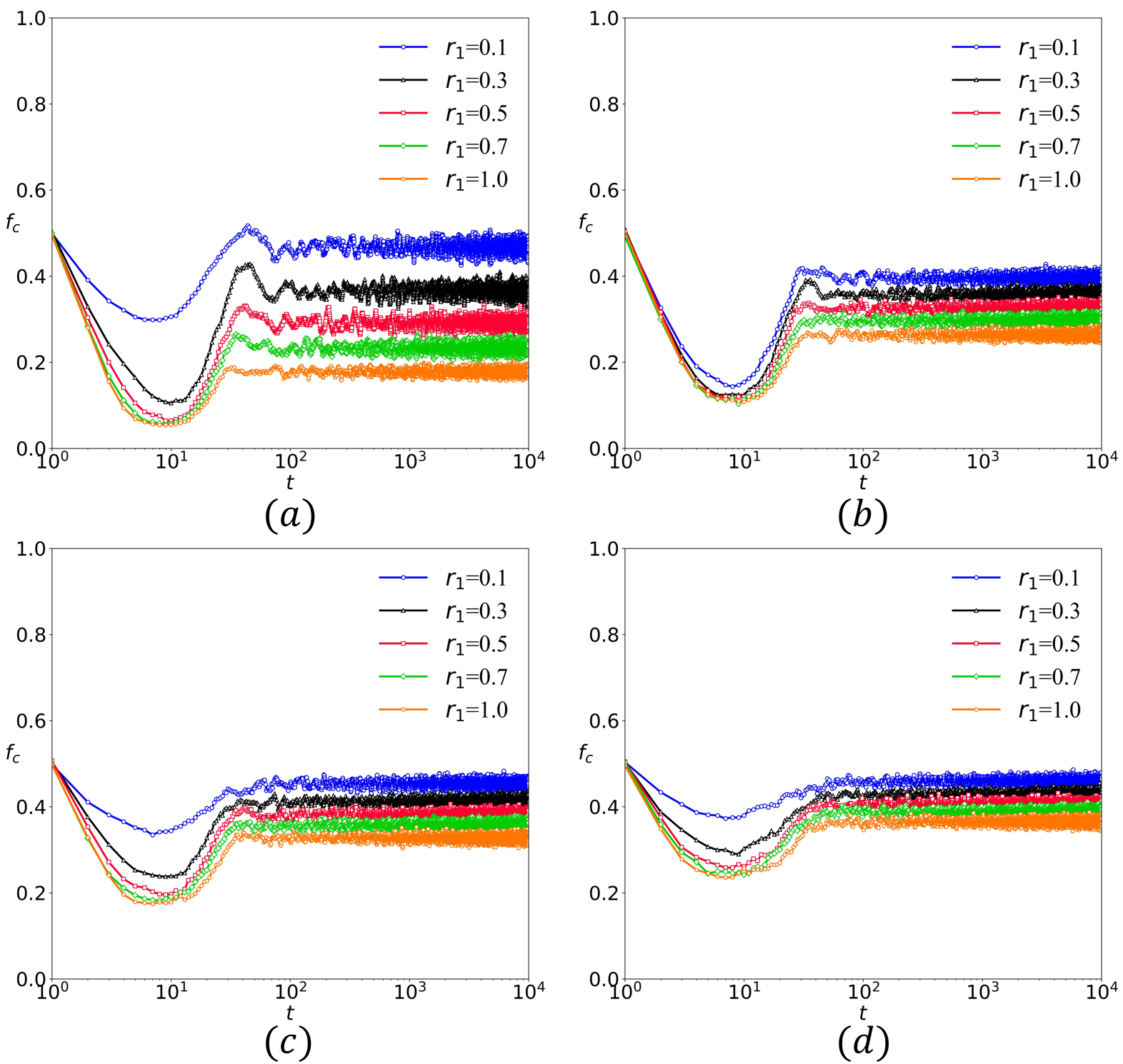}    
    \caption{\textbf{Time evolution of cooperation under different $m$ values obtained for WS graph in PDG.} Panel~(a) to (d) shows the trajectories obtained at $m=0.1$, $0.2$, $0.3$, and $0.4$, respectively. Each curve corresponds to different $r_1$ values, as indicated in the legend. Even though the trend of each curve in the four pictures is similar, the stationary value of $f_C$ increases as $m$ increases and gradually approaches $0.5$. }
    \label{Fig.MUTA}
\end{figure}

To finish our study of PDG we last studied how the mutation rate $m$ affects the general cooperation level in our model with extended updating dynamics. To make the results comparable to previous cases we keep WS network as an interaction graph and survey the cooperation level at different $m$ values in the full range of $r_1$ parameter. Figure~\ref{Fig.MUTA} summarizes our findings. Here we present the time evolution of the cooperation level starting from a random initial state for $m=0.1$, $0.2$, $0.3$, and $0.4$. 

The typical dynamics signaling network reciprocity can be seen for all cases, but the minimum values are gradually lifted by increasing $m$. As expected, the largest difference in the stationary $f_C$ levels for different $r_1$ values can be seen for the smallest $m$ value and these differences become marginal for very high $m$ values in panel~(d).

By comparing the four panels, we observe that as $m$ increases, the $f_c$ value at steady state decreases for low $r_1$ values ($r_1=0.1$ or $0.3$). It becomes evident that when $m$ exceeds $0.2$, the reputation mechanism becomes significantly less effective, and $f_c$ approaches $0.5$ after $t=10^2$. Therefore, controlling the value of $m$ is crucial. An excessively high value of $m$ can greatly diminish the effectiveness of the reputation mechanism. 
In Fig.~\ref{Fig.MUTA}(d), in contrast to Fig.~\ref{Fig.MUTA}(a) or (c), the amplitude of fluctuations in the $f_C$ curve decreases. During the descending phase, with $r_1=1.0$, the curve's minimum value only reaches approximately $0.25$. Furthermore, the cooperation frequency at the equilibrium state reaches a notable value of $0.4$, which exhibits minimal deviation compared to the curve at $r_1=0.1$.

In sum, we can conclude that increasing $m$ reduces the influence of $r_1$ on the resulting cooperative level, hence weakening the positive consequence of the reputation mechanism. Nevertheless, the final cooperation level is still beyond the portion dictated by the n{\"a}ive estimation of $m$ even at very large $r_1$ values, hence the positive consequence of the extended dynamics in collaboration with the generalized fitness function can still be detected. 

\begin{figure}[htpb]
	\centering
	\subfigure[SL with IM]{
		\includegraphics[scale=0.15]{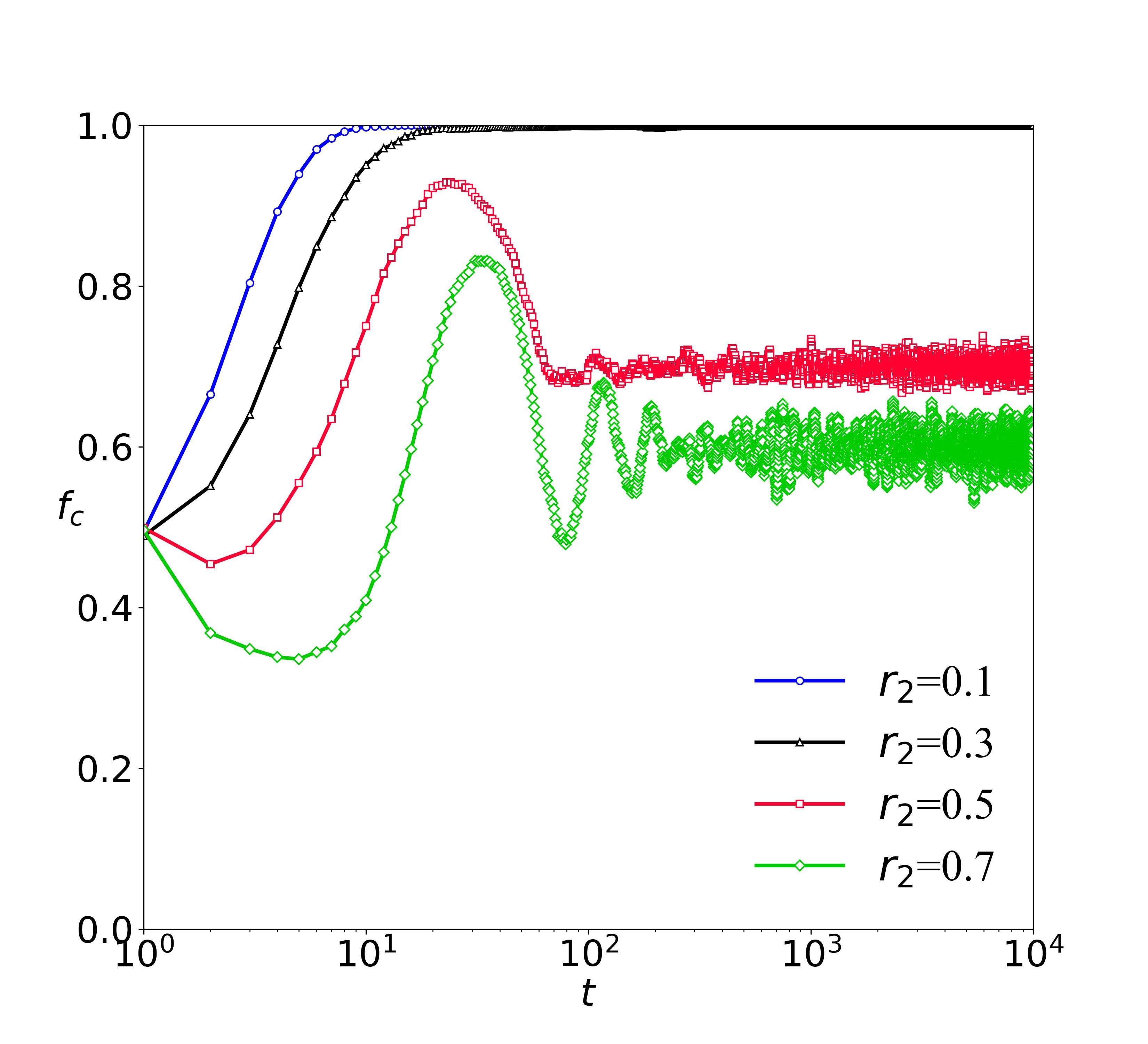}
		\label{e snapshots r=0.32}
  }
	\subfigure[WS with IM]{
		\includegraphics[scale=0.15]{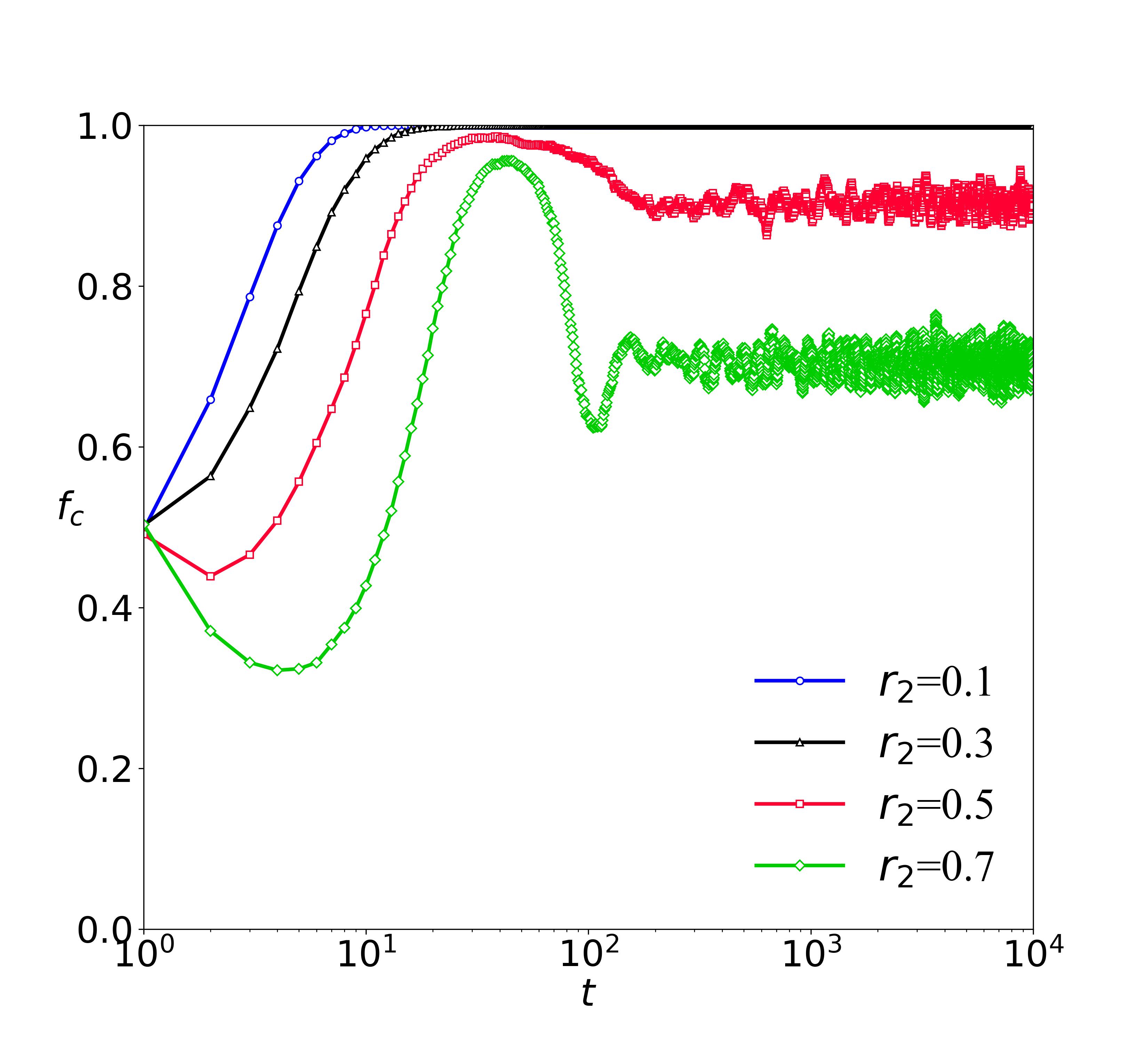}
		\label{e snapshots r=0.625}
  }
  \subfigure[BA with IM]{
		\includegraphics[scale=0.15]{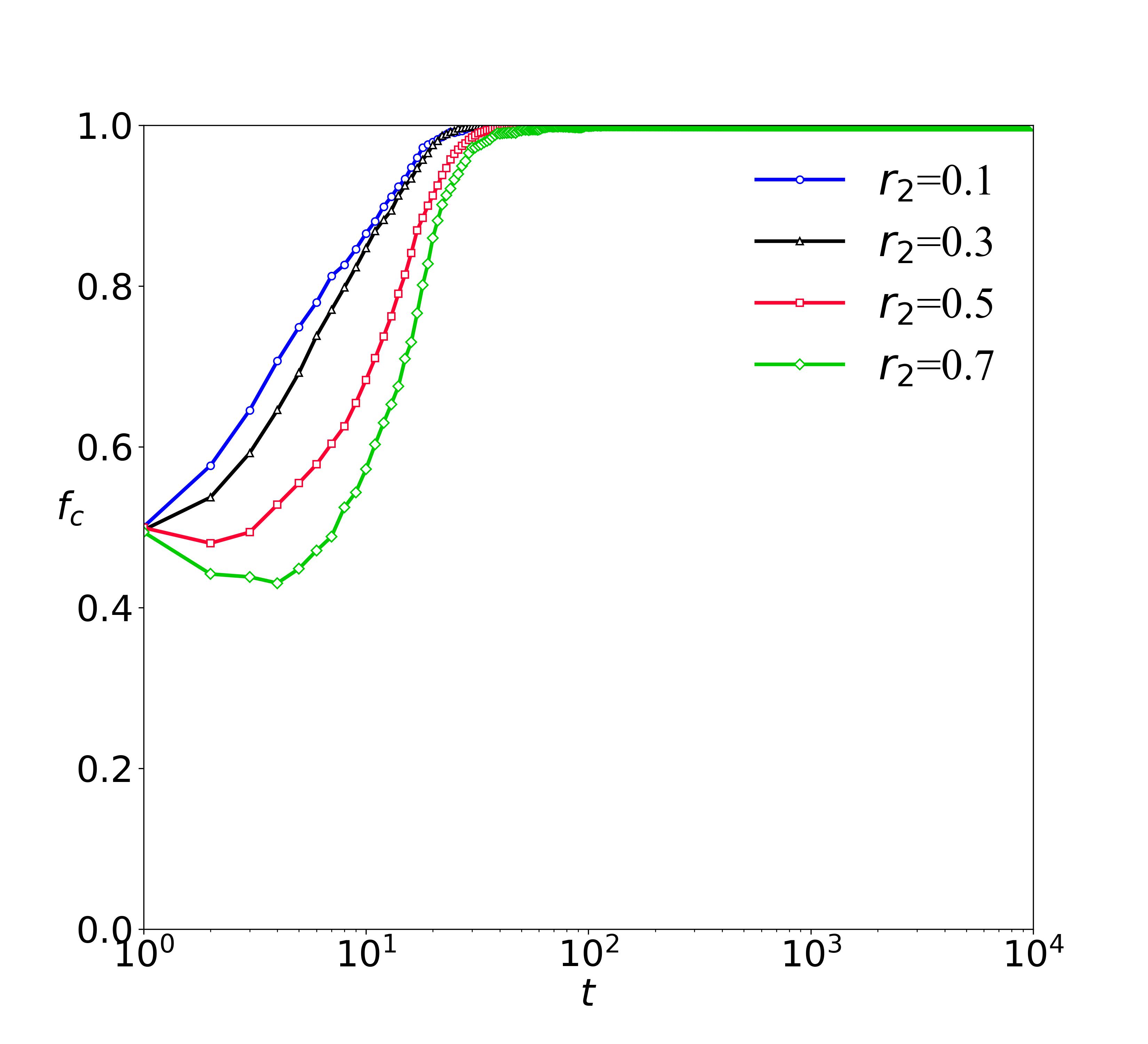}
		\label{e snapshots r=0.625}
  }
  
        \subfigure[SL with IM-MUTA]{
		\includegraphics[scale=0.15]{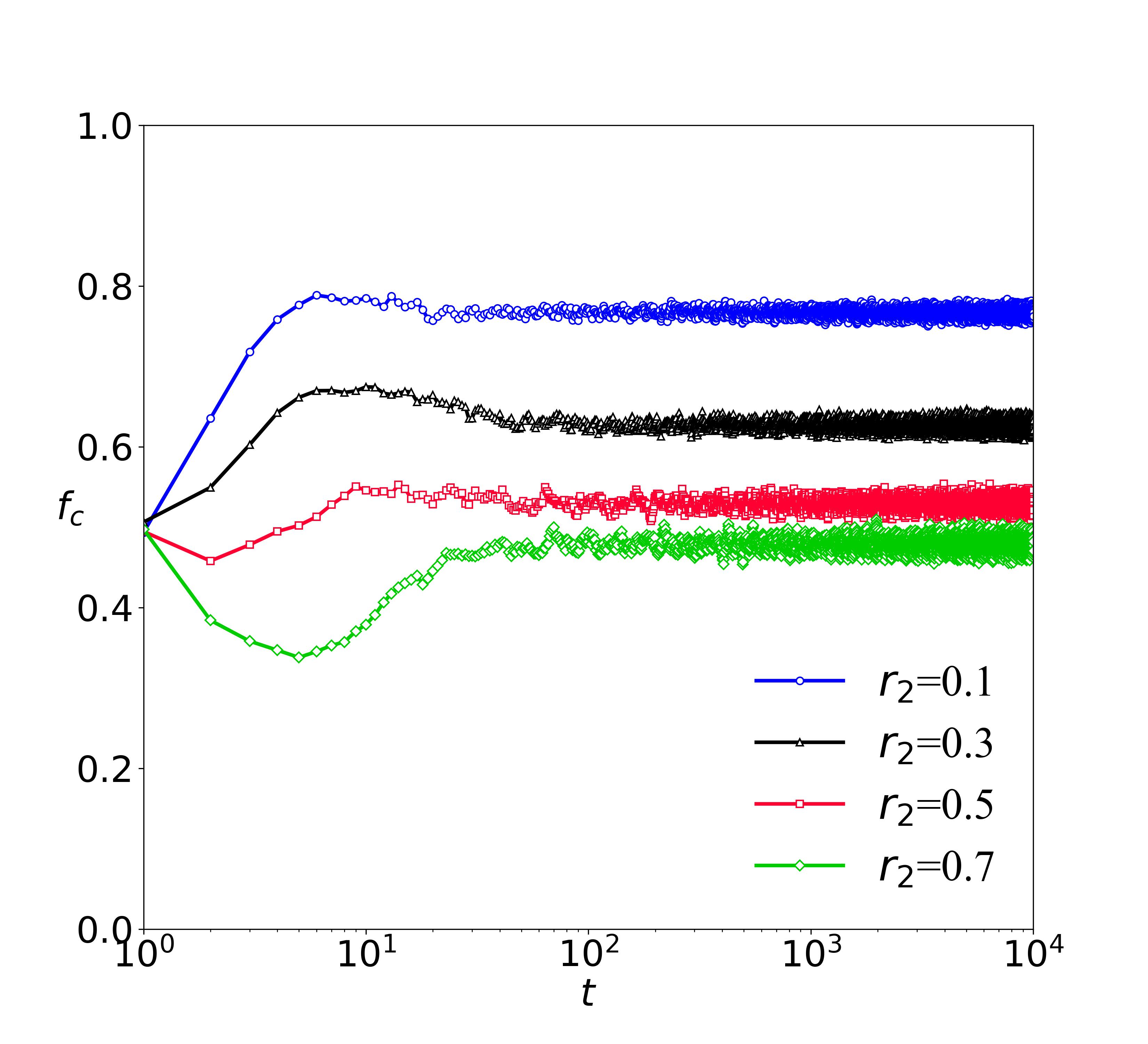}
		\label{K snapshots r=0.32}
  }
  	\subfigure[WS with IM-MUTA]{
		\includegraphics[scale=0.15]{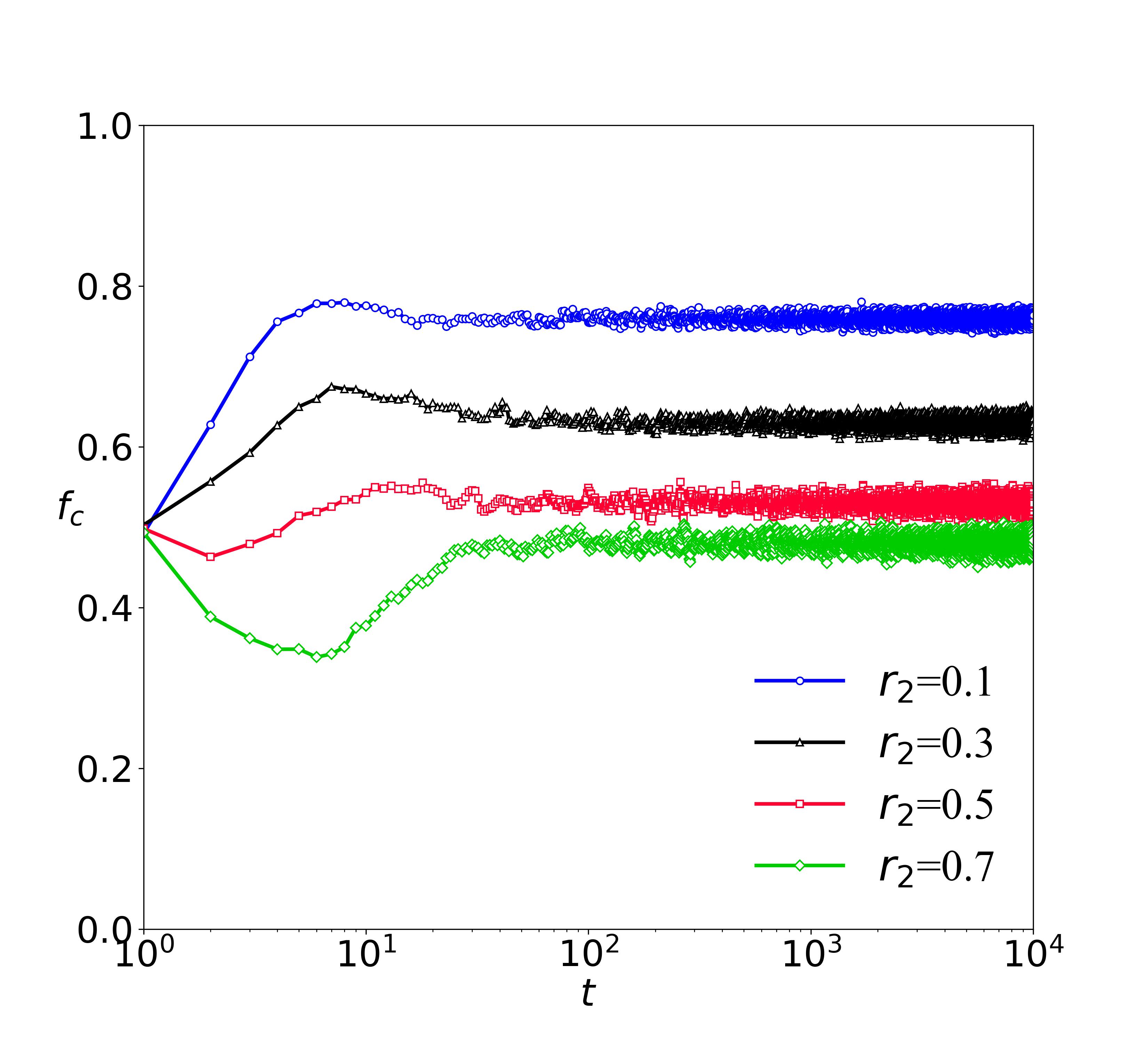}
		\label{K snapshots r=0.625}
  }
  \subfigure[BA with IM-MUTA]{
		\includegraphics[scale=0.15]{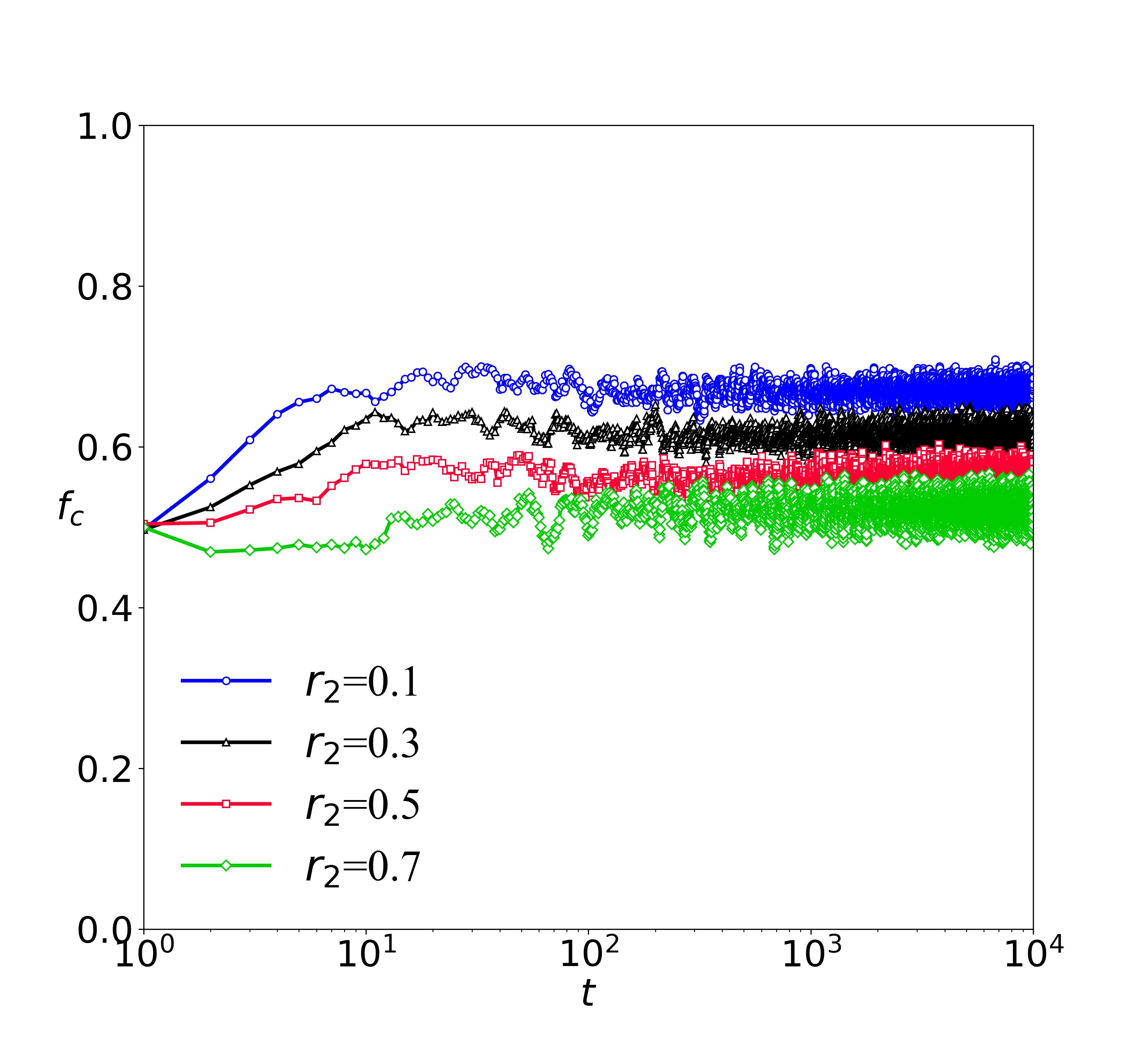}
		\label{K snapshots r=0.32}
  }
    %\captionsetup{justification=centerfirst}
	\caption{\textbf{The evolution of cooperation in SDG for different dynamics and topology.} Curves show the time evolution of $f_C$ starting from a random initial state for different $r_1$ values as indicated in the legend. The top row illustrates the evolution of the traditional model where players always follow imitation (IM) during strategy updates. Panel~(a) to (c) represent different interaction graphs, as shown in the labels. As a comparison, the bottom low depicts those cases where the extended imitation-mutation (IM-MUTA) strategy update is applied. Other parameters are $\alpha=0.05$ and $m=0.2$. Similarly to previous figures, we used a semi-log plot to stress the time-dependence more accurately.}
	\label{FIG:7}
\end{figure}

\subsection{Models in SDG game}

\begin{figure}
    \centering
    %\captionsetup{justification=centerfirst}
    \includegraphics[scale=.55]{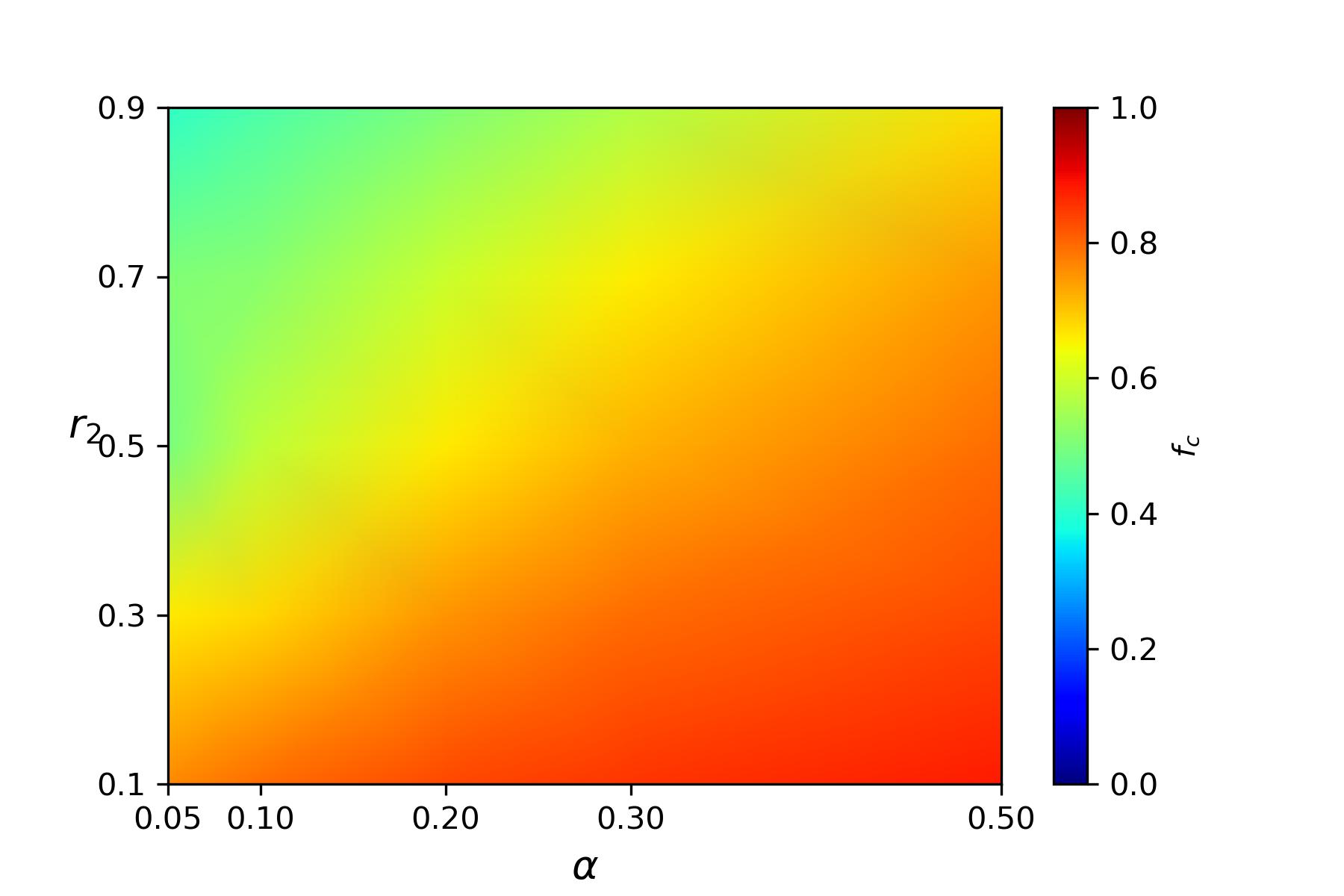}   
    \caption{\textbf{Stationary fraction of cooperators on $r_2 - \alpha$ parameter plane in SDG.} The results are obtained for WS interaction graph by using $m=0.2$. The impact of parameter $\alpha$ is clear, independently of $r_2$ value, the cooperation level can always be increased by using larger $\alpha$ steps during the reputation update. The largest cooperation level can be reached in the low $r_2$ - high $\alpha$ corner of the parameter plane.}
    \label{Fig.8}
\end{figure}

Finally, we complete our study by checking the newly introduced dynamics in SDG. Conceptually similar to the earlier discussed main section we now here apply three different types of interaction graphs, such as square lattice, WS small-world graph, and BA scale-free network. Besides, again for better comparison, we present results obtained by using the traditional and the extended dynamics. Our key observations are summarized in Fig.~\ref{FIG:7}. The top row denotes the time evolution in the traditional case when players always use imitation to update their strategies. The bottom row illustrates the system behavior when IM-MUTA update is used. As a general observation, the extended dynamics always provide a smaller cooperation level than the one we can reach by using solely imitation. The difference is especially shocking in the low $r_2$ interval where full cooperation can always be reached independently of the applied interaction graph. If, however, IM-MUTA is used, $f_C$ cannot exceed $0.8$. The smallest difference between the cooperation levels obtained for different dynamics can be detected for lattice structure in the large $r_2$ interval where network reciprocity has no relevance even in the traditional case~\cite{hauert_n04}. 

Evidently, the above-described observations were obtained at a specific $\alpha$ value, therefore we can ask how this parameter changes the final outcome. The answer is given in Fig.~\ref{Fig.8} where we plot the average cooperation level on the $r_2 - \alpha$ parameter plane. For a fair comparison obtained for PDG in Fig.~\ref{FIG:3}, we used WS interaction graph and $m=0.2$ mutation rate again. In stark contrast to the previous case, here the increase of $\alpha$ always improves cooperation independently of the applied $r_2$ value. Naturally, as previously, at higher temptation when we increase $r_2$ at a fixed $\alpha$ value, the cooperation level always decreases. The difference between the system behavior observed in PDG and SDG may be explained by the fact that SDG has different Nash-equilibrium hence the mixture of strategies provides the best strategy arrangement for an optimal global state even in the case when only imitation is allowed. Therefore the introduction of mutations does not really change the microscopic dynamics.

\section{ Conclusions and outlooks}

Reputation has been extensively examined in prior studies to assess its influence on the dynamics of cooperation. Nevertheless, existing literature predominantly focuses on the imitation of strategy update rules, often overlooking the presence of ``emergencies" or ``irrationalities'' in individual's decision-making processes \cite{tsetsos_pnas16,sircova_pone15}. This paper combines two aspects and introduces the concept of a mutation rate to individuals, which can be considered as an element of irrationality, allowing players to deviate from the pressure dictated by their neighborhood. The other key element of our model is the extended fitness function where reputation affects the likelihood of strategy imitation directly.

For a comprehensive study, we tested different types of interaction graphs, including lattices, random small-world, and scale-free topology. Furthermore, we also used alternative social dilemmas including prisoner's dilemma (PDG) and snow-drift game (SDG). In the former case, the usage of extended dynamics could be beneficial for cooperator strategy, particularly in high-temptation regions. This mechanism reduces the fluctuations in cooperation frequency and bolsters individuals' resistance to temptation. Conversely, in the case of SDG the usage of extended strategy updates is detrimental. As we noted, the difference can be explained in the diverse Nash-equilibrium characterizing these games.

The practical significance of these experimental results is noteworthy. When investigating the reputation growth coefficient, we have observed that a high reputation growth rate does not necessarily assist individuals in maintaining cooperation when faced with high-temptation situations. A rapid growth of reputation involves increased risks. When accumulated reputation becomes excessively high, any erroneous strategy choice can lead to a sharp decline in an individual's fitness, triggering widespread defection. This phenomenon mirrors real-life experience, where highly esteemed individuals, when embroiled in a scandal or controversy, incur significant costs and resource depletion, impacting both themselves and society at large. Unlike prior studies, networks employing our model may not achieve full cooperation. However, this outcome aligns more closely with the complexities of human society. 

However, the mutation mechanism employed in this study still has certain drawbacks. When the mutation rate $m$ is excessively high, the strategy updating of individuals tends to converge to a fixed probability. The mutation rate $m$ considered in this study acts as a global regulatory parameter that uniformly applies to all individuals in the network. Experimental results indicate that high values of $m$ inhibit the effectiveness of the reputation mechanism, transforming strategic selection into a pure probabilistic process. Hence, the cooperation frequency curve during the evolutionary process stabilizes, rendering the experimental results less realistic. To better simulate individual strategic decision-making in real-world social contexts, an improvement to the model could involve assigning a unique mutation rate within the network. Also, generating a new mutation rate for each individual in each round is a good method. As a result, our model would achieve a higher level of realism, making it better suited for simulating and understanding the evolution of cooperation. Also, we could address this limitation by incorporating expectation dynamics and combining different strategy updating rules to simulate human autonomous reasoning. Ultimately, we hope that our research can contribute to the field of evolutionary game theory involving self-awareness, group psychology, reputation mechanisms, and other related factors, thereby enhancing the practical relevance of these topics.

\section*{Acknowledgements}
This research was supported by the National Nature Science Foundation of China (NSFC) under Grant No.62206230, the Humanities and Social Science Fund of Ministry of Education of the People's Republic of China under Grant 21YJCZH028, the Natural Science Foundation of Chongqing under Grant No. CSTB2023NSCQ-MSX0064, and the National Research, Development and Innovation Office (NKFIH) under Grant No. K142948.

\section*{Data availability}
Data is available upon reasonable request.

\section*{Conflicts of interest}
The authors declare that they have no conflict of interest concerning the publication of this paper.

	\bibliographystyle{elsarticle-num-names}
	%\bibliography{cas-refs}

\begin{thebibliography}{70}
	\expandafter\ifx\csname natexlab\endcsname\relax\def\natexlab#1{#1}\fi
	\providecommand{\url}[1]{\texttt{#1}}
	\providecommand{\href}[2]{#2}
	\providecommand{\path}[1]{#1}
	\providecommand{\DOIprefix}{doi:}
	\providecommand{\ArXivprefix}{arXiv:}
	\providecommand{\URLprefix}{URL: }
	\providecommand{\Pubmedprefix}{pmid:}
	\providecommand{\doi}[1]{\href{http://dx.doi.org/#1}{\path{#1}}}
	\providecommand{\Pubmed}[1]{\href{pmid:#1}{\path{#1}}}
	\providecommand{\bibinfo}[2]{#2}
	\ifx\xfnm\relax \def\xfnm[#1]{\unskip,\space#1}\fi
	%Type = Article
	\bibitem[{Perc et~al.(2017)Perc, Jordan, Rand, Wang, Boccaletti, and
		Szolnoki}]{perc_pr17}
	\bibinfo{author}{M.~Perc}, \bibinfo{author}{J.~J. Jordan},
	\bibinfo{author}{D.~G. Rand}, \bibinfo{author}{Z.~Wang},
	\bibinfo{author}{S.~Boccaletti}, \bibinfo{author}{A.~Szolnoki},
	\newblock \bibinfo{title}{Statistical physics of human cooperation},
	\newblock \bibinfo{journal}{Phys. Rep.} \bibinfo{volume}{687}
	(\bibinfo{year}{2017}) \bibinfo{pages}{1--51}.
	%Type = Book
	\bibitem[{Nowak(2006)}]{nowak_06}
	\bibinfo{author}{M.~A. Nowak}, \bibinfo{title}{Evolutionary Dynamics},
	\bibinfo{publisher}{Harvard University Press}, \bibinfo{address}{Cambridge,
		MA}, \bibinfo{year}{2006}.
	%Type = Article
	\bibitem[{Kreft(2004)}]{kreft_mb04}
	\bibinfo{author}{J.-U. Kreft},
	\newblock \bibinfo{title}{Biofilms promote altruism},
	\newblock \bibinfo{journal}{Microbiology} \bibinfo{volume}{150}
	(\bibinfo{year}{2004}) \bibinfo{pages}{2751--2760}.
	%Type = Article
	\bibitem[{Chica et~al.(2019)Chica, Chiong, Adam, and Teubner}]{chica_srep19}
	\bibinfo{author}{M.~Chica}, \bibinfo{author}{R.~Chiong},
	\bibinfo{author}{M.~T.~P. Adam}, \bibinfo{author}{T.~Teubner},
	\newblock \bibinfo{title}{An evolutionary game model with punishment and
		protection to promote trust in the sharing economy},
	\newblock \bibinfo{journal}{Sci. Rep.} \bibinfo{volume}{9}
	(\bibinfo{year}{2019}) \bibinfo{pages}{19789}.
	%Type = Book
	\bibitem[{Axelrod(1984)}]{axelrod_84}
	\bibinfo{author}{R.~Axelrod}, \bibinfo{title}{The Evolution of Cooperation},
	\bibinfo{publisher}{Basic Books}, \bibinfo{address}{New York},
	\bibinfo{year}{1984}.
	%Type = Article
	\bibitem[{Wilkinson(1984)}]{wilkinson_n84}
	\bibinfo{author}{G.~S. Wilkinson},
	\newblock \bibinfo{title}{Reciprocal food sharing in the vampire bat},
	\newblock \bibinfo{journal}{Nature} \bibinfo{volume}{308}
	(\bibinfo{year}{1984}) \bibinfo{pages}{181--184}.
	%Type = Article
	\bibitem[{Datta and Gore(2014)}]{datta_cb14}
	\bibinfo{author}{M.~S. Datta}, \bibinfo{author}{J.~Gore},
	\newblock \bibinfo{title}{Evolution: ‘snowed’ in with the enemy},
	\newblock \bibinfo{journal}{Current Biology} \bibinfo{volume}{24}
	(\bibinfo{year}{2014}) \bibinfo{pages}{R34}.
	%Type = Article
	\bibitem[{Milinski et~al.(2008)Milinski, Sommerfeld, Krambeck, Reed, and
		Marotzke}]{milinski_pnas08}
	\bibinfo{author}{M.~Milinski}, \bibinfo{author}{R.~D. Sommerfeld},
	\bibinfo{author}{H.-J. Krambeck}, \bibinfo{author}{F.~A. Reed},
	\bibinfo{author}{J.~Marotzke},
	\newblock \bibinfo{title}{The collective-risk social dilemma and the prevention
		of simulated dangerous climate change},
	\newblock \bibinfo{journal}{Proc. Natl. Acad. Sci. U.S.A.}
	\bibinfo{volume}{105} (\bibinfo{year}{2008}) \bibinfo{pages}{2291--2294}.
	%Type = Article
	\bibitem[{Szolnoki(2014)}]{szolnoki_plrev14}
	\bibinfo{author}{A.~Szolnoki},
	\newblock \bibinfo{title}{The power of games: Comment on "climate change
		governance, cooperation and self-organization" by pacheco, vasconcelos and
		santos},
	\newblock \bibinfo{journal}{Phys. Life Rev.} \bibinfo{volume}{11}
	(\bibinfo{year}{2014}) \bibinfo{pages}{589--590}.
	%Type = Article
	\bibitem[{He et~al.(2023)He, Wang, Yu, Chen, Xu, and Dai}]{he_jl_pla23}
	\bibinfo{author}{J.~He}, \bibinfo{author}{J.~Wang}, \bibinfo{author}{F.~Yu},
	\bibinfo{author}{W.~Chen}, \bibinfo{author}{W.~Xu}, \bibinfo{author}{W.~Dai},
	\newblock \bibinfo{title}{Persistence-dependent dynamic interactive environment
		enhances cooperation},
	\newblock \bibinfo{journal}{Phys. Lett. A} \bibinfo{volume}{469}
	(\bibinfo{year}{2023}) \bibinfo{pages}{128748}.
	%Type = Article
	\bibitem[{Chen and Szolnoki(2018)}]{chen_xj_pcb18}
	\bibinfo{author}{X.~Chen}, \bibinfo{author}{A.~Szolnoki},
	\newblock \bibinfo{title}{Punishment and inspection for governing the commons
		in a feedback-evolving game},
	\newblock \bibinfo{journal}{PLoS Comput. Biol.} \bibinfo{volume}{14}
	(\bibinfo{year}{2018}) \bibinfo{pages}{e1006347}.
	%Type = Article
	\bibitem[{Axelrod and Hamilton(1981)}]{axelrod_s81}
	\bibinfo{author}{R.~Axelrod}, \bibinfo{author}{W.~D. Hamilton},
	\newblock \bibinfo{title}{The evolution of cooperation},
	\newblock \bibinfo{journal}{Science} \bibinfo{volume}{211}
	(\bibinfo{year}{1981}) \bibinfo{pages}{1390--1396}.
	%Type = Book
	\bibitem[{Maynard~Smith(1982)}]{maynard_82}
	\bibinfo{author}{J.~Maynard~Smith}, \bibinfo{title}{Evolution and the Theory of
		Games}, \bibinfo{publisher}{Cambridge University Press},
	\bibinfo{address}{Cambridge, U.K.}, \bibinfo{year}{1982}.
	%Type = Book
	\bibitem[{Sigmund(2010)}]{sigmund_10}
	\bibinfo{author}{K.~Sigmund}, \bibinfo{title}{The Calculus of Selfishness},
	\bibinfo{publisher}{Princeton University Press}, \bibinfo{address}{Princeton,
		NJ}, \bibinfo{year}{2010}.
	%Type = Book
	\bibitem[{Rapoport and Chammah(1970)}]{rapoport_70}
	\bibinfo{author}{A.~Rapoport}, \bibinfo{author}{A.~M. Chammah},
	\bibinfo{title}{Prisoner's Dilemma: A Study in Conflict and Cooperation},
	\bibinfo{publisher}{University of Michigan Press},
	\bibinfo{address}{Michigan}, \bibinfo{year}{1970}.
	%Type = Article
	\bibitem[{Zhang et~al.(2024)Zhang, Hao, Qian, Wu, Guo, and
		Ling}]{zhang_y_amc24}
	\bibinfo{author}{Y.~Zhang}, \bibinfo{author}{Q.-Y. Hao}, \bibinfo{author}{J.-L.
		Qian}, \bibinfo{author}{C.-Y. Wu}, \bibinfo{author}{N.~Guo},
	\bibinfo{author}{X.~Ling},
	\newblock \bibinfo{title}{The cooperative evolution in the spatial prisoner’s
		dilemma game with the local loyalty of two-strategy},
	\newblock \bibinfo{journal}{Appl. Math. Comput.} \bibinfo{volume}{466}
	(\bibinfo{year}{2024}) \bibinfo{pages}{128484}.
	%Type = Article
	\bibitem[{Yao et~al.(2023)Yao, Zeng, Pi, and Feng}]{YAO.2}
	\bibinfo{author}{Y.~Yao}, \bibinfo{author}{Z.~Zeng}, \bibinfo{author}{B.~Pi},
	\bibinfo{author}{M.~Feng},
	\newblock \bibinfo{title}{Inhibition and activation of interactions in
		networked weak prisoner’s dilemma},
	\newblock \bibinfo{journal}{Chaos} \bibinfo{volume}{33} (\bibinfo{year}{2023})
	\bibinfo{pages}{063124}.
	%Type = Article
	\bibitem[{Kojo and Sakiyama(2024)}]{kojo_csf24}
	\bibinfo{author}{K.~Kojo}, \bibinfo{author}{T.~Sakiyama},
	\newblock \bibinfo{title}{Restructuring of neighborhood definition based on
		strategies will enhance the cooperation in a spatial prisoner's dilemma},
	\newblock \bibinfo{journal}{Chaos, Solit. and Fract.} \bibinfo{volume}{179}
	(\bibinfo{year}{2024}) \bibinfo{pages}{114404}.
	%Type = Article
	\bibitem[{Mao et~al.(2021)Mao, Rong, and Wu}]{Rong.1}
	\bibinfo{author}{Y.~Mao}, \bibinfo{author}{Z.~Rong}, \bibinfo{author}{Z.~Wu},
	\newblock \bibinfo{title}{Effect of collective influence on the evolution of
		cooperation in evolutionary prisoner’s dilemma games},
	\newblock \bibinfo{journal}{Appl. Math. Comput.} \bibinfo{volume}{392}
	(\bibinfo{year}{2021}) \bibinfo{pages}{125679}.
	%Type = Article
	\bibitem[{Li et~al.(2021)Li, Mao, Wei, and Cong}]{li_k_csf21}
	\bibinfo{author}{K.~Li}, \bibinfo{author}{Y.~Mao}, \bibinfo{author}{Z.~Wei},
	\bibinfo{author}{R.~Cong},
	\newblock \bibinfo{title}{Pool-rewarding in n-person snowdrift game},
	\newblock \bibinfo{journal}{Chaos, Solit. and Fract.} \bibinfo{volume}{143}
	(\bibinfo{year}{2021}) \bibinfo{pages}{110591}.
	%Type = Article
	\bibitem[{Feng et~al.(2023)Feng, Han, Li, Wu, and Kurths}]{feng_my_csf23}
	\bibinfo{author}{M.~Feng}, \bibinfo{author}{S.~Han}, \bibinfo{author}{Q.~Li},
	\bibinfo{author}{J.~Wu}, \bibinfo{author}{J.~Kurths},
	\newblock \bibinfo{title}{Harmful strong agents and asymmetric interaction can
		promote the frequency of cooperation in the snowdrift game},
	\newblock \bibinfo{journal}{Chaos, Solit. and Fract.} \bibinfo{volume}{175}
	(\bibinfo{year}{2023}) \bibinfo{pages}{114068}.
	%Type = Article
	\bibitem[{Zeng et~al.(2022)Zeng, Li, and Feng}]{Zeng}
	\bibinfo{author}{Z.~Zeng}, \bibinfo{author}{Q.~Li}, \bibinfo{author}{M.~Feng},
	\newblock \bibinfo{title}{Spatial evolution of cooperation with variable
		payoffs},
	\newblock \bibinfo{journal}{Chaos} \bibinfo{volume}{32} (\bibinfo{year}{2022})
	\bibinfo{pages}{073118}.
	%Type = Book
	\bibitem[{Skyrms(2004)}]{skyrms_04}
	\bibinfo{author}{B.~Skyrms}, \bibinfo{title}{Stag-Hunt Game and the Evolution
		of Social Structure}, \bibinfo{publisher}{Cambridge University Press},
	\bibinfo{address}{Cambridge, U.K.}, \bibinfo{year}{2004}.
	%Type = Article
	\bibitem[{Deng and Zhang(2022)}]{deng_ys_epjb22}
	\bibinfo{author}{Y.~Deng}, \bibinfo{author}{J.~Zhang},
	\newblock \bibinfo{title}{The choice-decision based on memory and payoff favors
		cooperation in stag hunt game on interdependent networks},
	\newblock \bibinfo{journal}{Eur. Phys. J. B} \bibinfo{volume}{95}
	(\bibinfo{year}{2022}) \bibinfo{pages}{29}.
	%Type = Article
	\bibitem[{Yao et~al.(2023)Yao, Pi, Zeng, and Feng}]{YAO.1}
	\bibinfo{author}{Y.~Yao}, \bibinfo{author}{B.~Pi}, \bibinfo{author}{Z.~Zeng},
	\bibinfo{author}{M.~Feng},
	\newblock \bibinfo{title}{Protection and improvement of indirect identity
		cognition on the spatial evolution of cooperation},
	\newblock \bibinfo{journal}{Physica A} \bibinfo{volume}{620}
	(\bibinfo{year}{2023}) \bibinfo{pages}{128791}.
	%Type = Article
	\bibitem[{Szolnoki and Perc(2010)}]{szolnoki_epl10}
	\bibinfo{author}{A.~Szolnoki}, \bibinfo{author}{M.~Perc},
	\newblock \bibinfo{title}{Reward and cooperation in the spatial public goods
		game},
	\newblock \bibinfo{journal}{EPL} \bibinfo{volume}{92} (\bibinfo{year}{2010})
	\bibinfo{pages}{38003}.
	%Type = Article
	\bibitem[{Szolnoki and Perc(2009)}]{szolnoki_epl09}
	\bibinfo{author}{A.~Szolnoki}, \bibinfo{author}{M.~Perc},
	\newblock \bibinfo{title}{Resolving social dilemmas on evolving random
		networks},
	\newblock \bibinfo{journal}{EPL} \bibinfo{volume}{86} (\bibinfo{year}{2009})
	\bibinfo{pages}{30007}.
	%Type = Article
	\bibitem[{Dur{\'a}n and Mulet(2005)}]{duran_pd05}
	\bibinfo{author}{O.~Dur{\'a}n}, \bibinfo{author}{R.~Mulet},
	\newblock \bibinfo{title}{Evolutionary prisoner's dilemma in random graphs},
	\newblock \bibinfo{journal}{Physica D} \bibinfo{volume}{208}
	(\bibinfo{year}{2005}) \bibinfo{pages}{257--265}.
	%Type = Article
	\bibitem[{Vukov et~al.(2006)Vukov, Szab{\'o}, and Szolnoki}]{vukov_pre06}
	\bibinfo{author}{J.~Vukov}, \bibinfo{author}{G.~Szab{\'o}},
	\bibinfo{author}{A.~Szolnoki},
	\newblock \bibinfo{title}{Cooperation in the noisy case: Prisoner's dilemma
		game on two types of regular random graphs},
	\newblock \bibinfo{journal}{Phys. Rev. E} \bibinfo{volume}{73}
	(\bibinfo{year}{2006}) \bibinfo{pages}{067103}.
	%Type = Article
	\bibitem[{Kim et~al.(2002)Kim, Trusina, Holme, Minnhagen, Chung, and
		Choi}]{kim_bj_pre02}
	\bibinfo{author}{B.~J. Kim}, \bibinfo{author}{A.~Trusina},
	\bibinfo{author}{P.~Holme}, \bibinfo{author}{P.~Minnhagen},
	\bibinfo{author}{J.~S. Chung}, \bibinfo{author}{M.~Y. Choi},
	\newblock \bibinfo{title}{Dynamic instabilities induced by asymmetric
		influence: Prisoner's dilemma game in small-world networks},
	\newblock \bibinfo{journal}{Phys. Rev. E} \bibinfo{volume}{66}
	(\bibinfo{year}{2002}) \bibinfo{pages}{021907}.
	%Type = Article
	\bibitem[{Wu et~al.(2005)Wu, Xu, Chen, and Wang}]{wu_zx_pre05}
	\bibinfo{author}{Z.-X. Wu}, \bibinfo{author}{X.-J. Xu},
	\bibinfo{author}{Y.~Chen}, \bibinfo{author}{Y.-H. Wang},
	\newblock \bibinfo{title}{Spatial prisoner's dilemma game with volunteering in
		\protect{Newman-Watts} small-world networks},
	\newblock \bibinfo{journal}{Phys. Rev. E} \bibinfo{volume}{71}
	(\bibinfo{year}{2005}) \bibinfo{pages}{037103}.
	%Type = Article
	\bibitem[{Bin et~al.(2022)Bin, Li, and Feng}]{bin_pa22}
	\bibinfo{author}{P.~Bin}, \bibinfo{author}{Y.~Li}, \bibinfo{author}{M.~Feng},
	\newblock \bibinfo{title}{An evolutionary game with conformists and profiteers
		regarding the memory mechanism},
	\newblock \bibinfo{journal}{Physica A} \bibinfo{volume}{597}
	(\bibinfo{year}{2022}) \bibinfo{pages}{127297}.
	%Type = Article
	\bibitem[{Santos and Pacheco(2005)}]{santos_prl05}
	\bibinfo{author}{F.~C. Santos}, \bibinfo{author}{J.~M. Pacheco},
	\newblock \bibinfo{title}{Scale-free networks provide a unifying framework for
		the emergence of cooperation},
	\newblock \bibinfo{journal}{Phys. Rev. Lett.} \bibinfo{volume}{95}
	(\bibinfo{year}{2005}) \bibinfo{pages}{098104}.
	%Type = Article
	\bibitem[{Szolnoki et~al.(2008)Szolnoki, Perc, and Danku}]{szolnoki_pa08}
	\bibinfo{author}{A.~Szolnoki}, \bibinfo{author}{M.~Perc},
	\bibinfo{author}{Z.~Danku},
	\newblock \bibinfo{title}{Towards effective payoffs in the prisoner's dilemma
		game on scale-free networks},
	\newblock \bibinfo{journal}{Physica A} \bibinfo{volume}{387}
	(\bibinfo{year}{2008}) \bibinfo{pages}{2075--2082}.
	%Type = Article
	\bibitem[{Wang et~al.(2012)Wang, Szolnoki, and Perc}]{wang_z_epl12}
	\bibinfo{author}{Z.~Wang}, \bibinfo{author}{A.~Szolnoki},
	\bibinfo{author}{M.~Perc},
	\newblock \bibinfo{title}{Evolution of public cooperation on interdependent
		networks: The impact of biased utility functions},
	\newblock \bibinfo{journal}{EPL} \bibinfo{volume}{97} (\bibinfo{year}{2012})
	\bibinfo{pages}{48001}.
	%Type = Article
	\bibitem[{Li et~al.(2022)Li, Zhao, and Feng}]{li_q_e22}
	\bibinfo{author}{Q.~Li}, \bibinfo{author}{G.~Zhao}, \bibinfo{author}{M.~Feng},
	\newblock \bibinfo{title}{Prisoner's dilemma game with cooperation-defection
		dominance strategies on correlational multilayer networks},
	\newblock \bibinfo{journal}{Entropy} \bibinfo{volume}{24}
	(\bibinfo{year}{2022}) \bibinfo{pages}{822}.
	%Type = Article
	\bibitem[{Wang et~al.(2013)Wang, Szolnoki, and Perc}]{wang_z_srep13}
	\bibinfo{author}{Z.~Wang}, \bibinfo{author}{A.~Szolnoki},
	\bibinfo{author}{M.~Perc},
	\newblock \bibinfo{title}{Interdependent network reciprocity in evolutionary
		games},
	\newblock \bibinfo{journal}{Sci. Rep.} \bibinfo{volume}{3}
	(\bibinfo{year}{2013}) \bibinfo{pages}{1183}.
	%Type = Article
	\bibitem[{Mao et~al.(2023)Mao, Rong, Xu, and Han}]{Rong.2}
	\bibinfo{author}{Y.~Mao}, \bibinfo{author}{Z.~Rong}, \bibinfo{author}{X.~Xu},
	\bibinfo{author}{Z.~Han},
	\newblock \bibinfo{title}{Influence of diverse timescales on the evolution of
		cooperation in a double-layer lattice},
	\newblock \bibinfo{journal}{Front. Phys.} \bibinfo{volume}{11}
	(\bibinfo{year}{2023}) \bibinfo{pages}{1272395}.
	%Type = Article
	\bibitem[{Yang and Yang(2019)}]{yang_hx_pa19}
	\bibinfo{author}{H.-X. Yang}, \bibinfo{author}{J.~Yang},
	\newblock \bibinfo{title}{Reputation-based investment strategy promotes
		cooperation in public goods games},
	\newblock \bibinfo{journal}{Physica A} \bibinfo{volume}{523}
	(\bibinfo{year}{2019}) \bibinfo{pages}{886--893}.
	%Type = Article
	\bibitem[{Quan et~al.(2021)Quan, Tang, and Wang}]{quan_j_pa21}
	\bibinfo{author}{J.~Quan}, \bibinfo{author}{C.~Tang},
	\bibinfo{author}{X.~Wang},
	\newblock \bibinfo{title}{Reputation-based discount effect in imitation on the
		evolution of cooperation in spatial public goods games},
	\newblock \bibinfo{journal}{Physica A} \bibinfo{volume}{563}
	(\bibinfo{year}{2021}) \bibinfo{pages}{125488}.
	%Type = Article
	\bibitem[{Helbing et~al.(2010)Helbing, Szolnoki, Perc, and
		Szab{\'o}}]{helbing_ploscb10}
	\bibinfo{author}{D.~Helbing}, \bibinfo{author}{A.~Szolnoki},
	\bibinfo{author}{M.~Perc}, \bibinfo{author}{G.~Szab{\'o}},
	\newblock \bibinfo{title}{Evolutionary establishment of moral and double moral
		standards through spatial interactions},
	\newblock \bibinfo{journal}{PLoS Comput. Biol.} \bibinfo{volume}{6}
	(\bibinfo{year}{2010}) \bibinfo{pages}{e1000758}.
	%Type = Article
	\bibitem[{Brandt et~al.(2003)Brandt, Hauert, and Sigmund}]{brandt_prsb03}
	\bibinfo{author}{H.~Brandt}, \bibinfo{author}{C.~Hauert},
	\bibinfo{author}{K.~Sigmund},
	\newblock \bibinfo{title}{Punishing and reputation in spatial public goods
		games},
	\newblock \bibinfo{journal}{Proc. R. Soc. Lond. Ser B} \bibinfo{volume}{270}
	(\bibinfo{year}{2003}) \bibinfo{pages}{1099--1104}.
	%Type = Article
	\bibitem[{Lee et~al.(2022)Lee, Cleveland, and Szolnoki}]{lee_hw_amc22}
	\bibinfo{author}{H.-W. Lee}, \bibinfo{author}{C.~Cleveland},
	\bibinfo{author}{A.~Szolnoki},
	\newblock \bibinfo{title}{Mercenary punishment in structured populations},
	\newblock \bibinfo{journal}{Appl. Math. Comput.} \bibinfo{volume}{417}
	(\bibinfo{year}{2022}) \bibinfo{pages}{126797}.
	%Type = Article
	\bibitem[{Liu and Chen(2022)}]{liu_lj_rspa22}
	\bibinfo{author}{L.~Liu}, \bibinfo{author}{X.~Chen},
	\newblock \bibinfo{title}{Indirect exclusion can promote cooperation in
		repeated group interactions},
	\newblock \bibinfo{journal}{Proc. R. Soc. A} \bibinfo{volume}{478}
	(\bibinfo{year}{2022}) \bibinfo{pages}{20220290}.
	%Type = Article
	\bibitem[{Szolnoki and Chen(2017)}]{szolnoki_pre17}
	\bibinfo{author}{A.~Szolnoki}, \bibinfo{author}{X.~Chen},
	\newblock \bibinfo{title}{Alliance formation with exclusion in the spatial
		public goods game},
	\newblock \bibinfo{journal}{Phys. Rev. E} \bibinfo{volume}{95}
	(\bibinfo{year}{2017}) \bibinfo{pages}{052316}.
	%Type = Article
	\bibitem[{Quan et~al.(2022)Quan, Guo, and Wang}]{quan_j_jsm22}
	\bibinfo{author}{J.~Quan}, \bibinfo{author}{H.~Guo}, \bibinfo{author}{X.~Wang},
	\newblock \bibinfo{title}{Impact of reputation-based switching strategy between
		punishment and social exclusion on the evolution of cooperation in the
		spatial public goods game},
	\newblock \bibinfo{journal}{J. Stat. Mech.} \bibinfo{volume}{2022}
	(\bibinfo{year}{2022}) \bibinfo{pages}{073402}.
	%Type = Article
	\bibitem[{Sun et~al.(2023)Sun, Han, Wang, Liu, and Shen}]{sun_xp_pla23}
	\bibinfo{author}{X.~Sun}, \bibinfo{author}{L.~Han}, \bibinfo{author}{M.~Wang},
	\bibinfo{author}{S.~Liu}, \bibinfo{author}{Y.~Shen},
	\newblock \bibinfo{title}{Social exclusion with antisocial punishment in
		spatial public goods game},
	\newblock \bibinfo{journal}{Phys. Lett. A} \bibinfo{volume}{474}
	(\bibinfo{year}{2023}) \bibinfo{pages}{128837}.
	%Type = Article
	\bibitem[{Hua and Liu(2024)}]{hua_sj_eswa24}
	\bibinfo{author}{S.~Hua}, \bibinfo{author}{L.~Liu},
	\newblock \bibinfo{title}{Coevolutionary dynamics of population and
		institutional rewards in public goods games},
	\newblock \bibinfo{journal}{Expert Systems With Applications}
	\bibinfo{volume}{237} (\bibinfo{year}{2024}) \bibinfo{pages}{121579}.
	%Type = Article
	\bibitem[{Zhang et~al.(2019)Zhang, Huang, Li, and Dai}]{zhang_lm_epl19}
	\bibinfo{author}{L.~Zhang}, \bibinfo{author}{C.~Huang},
	\bibinfo{author}{H.~Li}, \bibinfo{author}{Q.~Dai},
	\newblock \bibinfo{title}{Aspiration-dependent strategy persistence promotes
		cooperation in spatial prisoner's dilemma game},
	\newblock \bibinfo{journal}{EPL} \bibinfo{volume}{126} (\bibinfo{year}{2019})
	\bibinfo{pages}{18001}.
	%Type = Article
	\bibitem[{Liu et~al.(2019)Liu, Huang, Dai, and Li}]{liu_dn_pa19}
	\bibinfo{author}{D.~Liu}, \bibinfo{author}{C.~Huang}, \bibinfo{author}{Q.~Dai},
	\bibinfo{author}{H.~Li},
	\newblock \bibinfo{title}{Positive correlation between strategy persistence and
		teaching ability promotes cooperation in evolutionary prisoner's dilemma
		games},
	\newblock \bibinfo{journal}{Physica A} \bibinfo{volume}{520}
	(\bibinfo{year}{2019}) \bibinfo{pages}{267--274}.
	%Type = Article
	\bibitem[{Nowak and Sigmund(1993)}]{nowak_n93}
	\bibinfo{author}{M.~A. Nowak}, \bibinfo{author}{K.~Sigmund},
	\newblock \bibinfo{title}{A strategy of win-stay, lose-shift that outperforms
		tit-for-tat in the prisoner's dilemma game},
	\newblock \bibinfo{journal}{Nature} \bibinfo{volume}{364}
	(\bibinfo{year}{1993}) \bibinfo{pages}{56--58}.
	%Type = Article
	\bibitem[{Sasidevan and Sinha(2016)}]{sasidevan_srep16}
	\bibinfo{author}{V.~Sasidevan}, \bibinfo{author}{S.~Sinha},
	\newblock \bibinfo{title}{Co-action provides rational basis for the
		evolutionary success of \protect{Pavlovian} strategies},
	\newblock \bibinfo{journal}{Sci. Rep.} \bibinfo{volume}{6}
	(\bibinfo{year}{2016}) \bibinfo{pages}{30831}.
	%Type = Article
	\bibitem[{Fu and Yang(2018)}]{fu_mj_ijmpc18}
	\bibinfo{author}{M.-J. Fu}, \bibinfo{author}{H.-X. Yang},
	\newblock \bibinfo{title}{Stochastic win-stay-lose-learn promotes cooperation
		in the spatial public goods gam},
	\newblock \bibinfo{journal}{Int. J. Mod. Phys. C} \bibinfo{volume}{29}
	(\bibinfo{year}{2018}) \bibinfo{pages}{1850034}.
	%Type = Article
	\bibitem[{Su et~al.(2016)Su, Li, Zhou, and Wang}]{su_q_njp16}
	\bibinfo{author}{Q.~Su}, \bibinfo{author}{A.~Li}, \bibinfo{author}{L.~Zhou},
	\bibinfo{author}{L.~Wang},
	\newblock \bibinfo{title}{Interactive diversity promotes the evolution of
		cooperation in structured populations},
	\newblock \bibinfo{journal}{New J. Phys.} \bibinfo{volume}{18}
	(\bibinfo{year}{2016}) \bibinfo{pages}{103007}.
	%Type = Article
	\bibitem[{Feng et~al.(2024)Feng, Pi, Deng, and Kurths}]{M.Feng}
	\bibinfo{author}{M.~Feng}, \bibinfo{author}{B.~Pi}, \bibinfo{author}{L.~Deng},
	\bibinfo{author}{J.~Kurths},
	\newblock \bibinfo{title}{An evolutionary game with the game transitions based
		on the \protect{Markov} process},
	\newblock \bibinfo{journal}{IEEE Trans. Syst. Man Cybern.} \bibinfo{volume}{54}
	(\bibinfo{year}{2024}) \bibinfo{pages}{609--621}.
	%Type = Article
	\bibitem[{Wei et~al.(2021)Wei, Xu, Du, Yan, and Pei}]{wei_x_epjb21}
	\bibinfo{author}{X.~Wei}, \bibinfo{author}{P.~Xu}, \bibinfo{author}{S.~Du},
	\bibinfo{author}{G.~Yan}, \bibinfo{author}{H.~Pei},
	\newblock \bibinfo{title}{Reputational preference-based payoff punishment
		promotes cooperation in spatial social dilemmas},
	\newblock \bibinfo{journal}{Eur. Phys. J. B} \bibinfo{volume}{94}
	(\bibinfo{year}{2021}) \bibinfo{pages}{210}.
	%Type = Article
	\bibitem[{Bin and Yue(2023)}]{bin_l_amc23}
	\bibinfo{author}{L.~Bin}, \bibinfo{author}{W.~Yue},
	\newblock \bibinfo{title}{Co-evolution of reputation-based preference selection
		and resource allocation with multigame on interdependent networks},
	\newblock \bibinfo{journal}{Appl. Math. Comput.} \bibinfo{volume}{456}
	(\bibinfo{year}{2023}) \bibinfo{pages}{128128}.
	%Type = Article
	\bibitem[{Gruji{\'c} and Lenaerts(2022)}]{grujic_rsos22}
	\bibinfo{author}{J.~Gruji{\'c}}, \bibinfo{author}{T.~Lenaerts},
	\newblock \bibinfo{title}{Do people imitate when making decisions? evidence
		from a spatial prisoner’s dilemma experiment},
	\newblock \bibinfo{journal}{R. Soc. Open Sci.} \bibinfo{volume}{7}
	(\bibinfo{year}{2022}) \bibinfo{pages}{200618}.
	%Type = Article
	\bibitem[{Rizzolatti and Craighero(2004)}]{rizolatti_arn04}
	\bibinfo{author}{G.~Rizzolatti}, \bibinfo{author}{L.~Craighero},
	\newblock \bibinfo{title}{The mirror-neuron system},
	\newblock \bibinfo{journal}{Annu. Rev. Neurosci.} \bibinfo{volume}{27}
	(\bibinfo{year}{2004}) \bibinfo{pages}{169--92}.
	%Type = Article
	\bibitem[{Zhang et~al.(2022)Zhang, Wu, and Zhang}]{zhang_zp_pa22}
	\bibinfo{author}{Z.~Zhang}, \bibinfo{author}{Y.~Wu},
	\bibinfo{author}{S.~Zhang},
	\newblock \bibinfo{title}{Reputation-based asymmetric comparison of fitness
		promotes cooperation on complex networks},
	\newblock \bibinfo{journal}{Physica A} \bibinfo{volume}{608}
	(\bibinfo{year}{2022}) \bibinfo{pages}{128268}.
	%Type = Article
	\bibitem[{Vasconcelos et~al.(2015)Vasconcelos, Monteiro, and
		Kacelnik}]{irrational.1}
	\bibinfo{author}{M.~Vasconcelos}, \bibinfo{author}{T.~Monteiro},
	\bibinfo{author}{A.~Kacelnik},
	\newblock \bibinfo{title}{Irrational choice and the value of information},
	\newblock \bibinfo{journal}{Sci. Rep.} \bibinfo{volume}{5}
	(\bibinfo{year}{2015}) \bibinfo{pages}{13874}.
	%Type = Article
	\bibitem[{Opaluch and Segerson(1989)}]{irrational.2}
	\bibinfo{author}{J.~Opaluch}, \bibinfo{author}{K.~Segerson},
	\newblock \bibinfo{title}{Rational roots of “irrational” behavior: New
		theories of economic decision-making},
	\newblock \bibinfo{journal}{Northeast. J. Agric. Resour. Econ.}
	\bibinfo{volume}{18} (\bibinfo{year}{1989}) \bibinfo{pages}{81--95}.
	%Type = Article
	\bibitem[{Traulsen et~al.(2009)Traulsen, Hauert, De~Silva, Nowak, and
		Sigmund}]{traulsen_pnas09}
	\bibinfo{author}{A.~Traulsen}, \bibinfo{author}{C.~Hauert},
	\bibinfo{author}{H.~De~Silva}, \bibinfo{author}{M.~A. Nowak},
	\bibinfo{author}{K.~Sigmund},
	\newblock \bibinfo{title}{Exploration dynamics in evolutionary games},
	\newblock \bibinfo{journal}{Proc. Natl. Acad. Sci. USA} \bibinfo{volume}{106}
	(\bibinfo{year}{2009}) \bibinfo{pages}{709--712}.
	%Type = Article
	\bibitem[{Santos et~al.(2016)Santos, Pacheco, and Santos}]{santos_srep16}
	\bibinfo{author}{F.~P. Santos}, \bibinfo{author}{J.~M. Pacheco},
	\bibinfo{author}{F.~C. Santos},
	\newblock \bibinfo{title}{Evolution of cooperation under indirect reciprocity
		and arbitrary exploration rates},
	\newblock \bibinfo{journal}{Sci. Rep.} \bibinfo{volume}{6}
	(\bibinfo{year}{2016}) \bibinfo{pages}{37517}.
	%Type = Article
	\bibitem[{Erovenko et~al.(2019)Erovenko, Bauer, Broom, Pattni, and Rycht{\'
			a}{\v r}}]{erovenko_prsa19}
	\bibinfo{author}{I.~V. Erovenko}, \bibinfo{author}{J.~Bauer},
	\bibinfo{author}{M.~Broom}, \bibinfo{author}{K.~Pattni},
	\bibinfo{author}{J.~Rycht{\' a}{\v r}},
	\newblock \bibinfo{title}{The effect of network topology on optimal exploration
		strategies and the evolution of cooperation in a mobile population},
	\newblock \bibinfo{journal}{Proc. R. Soc. A} \bibinfo{volume}{475}
	(\bibinfo{year}{2019}) \bibinfo{pages}{20190399}.
	%Type = Article
	\bibitem[{Perc et~al.(2008)Perc, Szolnoki, and Szab{\'o}}]{perc_pre08b}
	\bibinfo{author}{M.~Perc}, \bibinfo{author}{A.~Szolnoki},
	\bibinfo{author}{G.~Szab{\'o}},
	\newblock \bibinfo{title}{Restricted connections among distinguished players
		support cooperation},
	\newblock \bibinfo{journal}{Phys. Rev. E} \bibinfo{volume}{78}
	(\bibinfo{year}{2008}) \bibinfo{pages}{066101}.
	%Type = Article
	\bibitem[{Szolnoki and Perc(2009)}]{szolnoki_epjb09}
	\bibinfo{author}{A.~Szolnoki}, \bibinfo{author}{M.~Perc},
	\newblock \bibinfo{title}{Promoting cooperation in social dilemmas via simple
		coevolutionary rules},
	\newblock \bibinfo{journal}{Eur. Phys. J. B} \bibinfo{volume}{67}
	(\bibinfo{year}{2009}) \bibinfo{pages}{337--344}.
	%Type = Article
	\bibitem[{Hauert and Doebeli(2004)}]{hauert_n04}
	\bibinfo{author}{C.~Hauert}, \bibinfo{author}{M.~Doebeli},
	\newblock \bibinfo{title}{Spatial structure often inhibits the evolution of
		cooperation in the snowdrift game},
	\newblock \bibinfo{journal}{Nature} \bibinfo{volume}{428}
	(\bibinfo{year}{2004}) \bibinfo{pages}{643--646}.
	%Type = Article
	\bibitem[{Tsetsos et~al.(2016)Tsetsos, Moran, Moreland, Chater, Usher, and
		Summerfield}]{tsetsos_pnas16}
	\bibinfo{author}{K.~Tsetsos}, \bibinfo{author}{R.~Moran},
	\bibinfo{author}{J.~Moreland}, \bibinfo{author}{N.~Chater},
	\bibinfo{author}{M.~Usher}, \bibinfo{author}{C.~Summerfield},
	\newblock \bibinfo{title}{Economic irrationality is optimal during noisy
		decision making},
	\newblock \bibinfo{journal}{Proc. Natl. Acad. Sci. U.S. A.}
	\bibinfo{volume}{113} (\bibinfo{year}{2016}) \bibinfo{pages}{3102--3107}.
	%Type = Article
	\bibitem[{Sircova et~al.(2015)Sircova, Karimi, Osin, Lee, Holme, and Str{\"
			o}mbom}]{sircova_pone15}
	\bibinfo{author}{A.~Sircova}, \bibinfo{author}{F.~Karimi},
	\bibinfo{author}{E.~N. Osin}, \bibinfo{author}{S.~Lee},
	\bibinfo{author}{P.~Holme}, \bibinfo{author}{D.~Str{\" o}mbom},
	\newblock \bibinfo{title}{Simulating irrational human behavior to prevent
		resource depletion},
	\newblock \bibinfo{journal}{PLoS ONE} \bibinfo{volume}{10}
	(\bibinfo{year}{2015}) \bibinfo{pages}{e0117612}.
	
\end{thebibliography}

\end{document}